\documentclass[aps,prx,twocolumn,superscriptaddress,showpacs,final,floatfix,longbibliography]{revtex4-2}

\usepackage[utf8]{inputenc} 
\usepackage[toc,page]{appendix}
\usepackage{amsfonts}
\usepackage[dvips]{graphicx}
\usepackage{color}
\usepackage{amsmath}
\usepackage{graphicx}
\usepackage{amssymb}
\usepackage{hyperref}
\usepackage{color}
\usepackage{mathrsfs}
\usepackage{isomath}
\usepackage{amsthm}
\usepackage{epstopdf}
\usepackage{txfonts}
\usepackage{dsfont}
\usepackage{ulem}
\usepackage{physics} % braket
\usepackage{tikz} 
\allowdisplaybreaks[4]
\usepackage{bbold}
\usepackage{mathtools}
\usepackage{cleveref}
\usepackage{subfigure}

\newcommand{\ignore}[1]{}

\newcommand{\de}{{\mathrm d}}

\renewcommand{\vec}[1]{\boldsymbol{#1}}

\newcommand{\hil}{\mathcal{H}}

\renewcommand{\tr}{\mathrm{tr}} %trace

\newcommand{\be}{\begin{equation}}
\newcommand{\ee}{\end{equation}}
\newcommand{\eea}{\end{eqnarray}}
\newcommand{\bea}{\begin{eqnarray}}
\renewcommand{\var}[1]{\ensuremath{(\Delta #1)^2}}
\newcommand{\av}[1]{\ensuremath{\langle{#1} \rangle}}

\renewcommand{\vec}[1]{\boldsymbol{#1}}

\newcommand{\id}{\mathbb{1}}
\newtheorem{corollary}{Corollary}
\newtheorem{observation}{Observation}
\newtheorem{lemma}{Lemma}

\usepackage{cleveref}
\crefname{equation}{Eq.}{Eqs.}
\crefname{figure}{Fig.}{Figs.}
\crefname{observation}{Obs.}{Obs.}
\crefname{corollary}{Corollary}{Corollaries}
\crefname{lemma}{Lemma}{Lemmata}
\crefname{proof}{Proof}{Proofs}
\creflabelformat{proof}{#2proof#3}
\crefname{remark}{Remark}{Remarks}
\crefname{prop}{Proposition}{Propositions}

\newcommand{\smcref}[1]{SM Sec.~\ref{#1}}

\begin{document}

\title{
Uncertainty relations between quantum Fisher information and entanglement monotones
}

\author{Shaowei Du$^{\ddagger}$}
\affiliation{State key Laboratory of Artificial Microstructure and Mesoscopic Physics, School of Physics, Frontiers Science Center for Nano-optoelectronics, Peking University, Beijing 100871, China}

\author{Shuheng Liu$^{\ddagger}$}
\email{liushuheng@pku.edu.cn}
\affiliation{State key Laboratory of Artificial Microstructure and Mesoscopic Physics, School of Physics, Frontiers Science Center for Nano-optoelectronics, Peking University, Beijing 100871, China}

\author{Matteo Fadel}%\orcid{0000-0003-3653-0030}
\affiliation{Department of Physics, ETH Z\"{u}rich, 8093 Z\"{u}rich, Switzerland}

\author{Giuseppe Vitagliano}%\orcidlink{0000-0002-5563-3222}
\affiliation{Vienna Center for Quantum Science and Technology, Atominstitut, TU Wien, 1020 Vienna, Austria}

\author{Qiongyi He}%\orcidlink{0000-0002-2408-4320}
\email{qiongyihe@pku.edu.cn}
\affiliation{State key Laboratory of Artificial Microstructure and Mesoscopic Physics, School of Physics, Frontiers Science Center for Nano-optoelectronics, Peking University, Beijing 100871, China}
\affiliation{Collaborative Innovation Center of Extreme Optics, Shanxi University, Taiyuan, Shanxi 030006, China}
\affiliation{Hefei National Laboratory, Hefei 230088, China}

\begin{abstract}
Entanglement is widely regarded as an essential resource for a number of tasks and can in some cases be quantified by figures of merit related to those tasks. In quantum metrology, this is showcased by the connections between the quantum Fisher information (QFI), providing a bound to the precision, and multipartite entanglement quantifiers such as the entanglement depth. However, a connection between the QFI and entanglement {\it monotones}, i.e., functions that do not increase under Local Operations and Classical Communications, has so far remained elusive. In this work, we fill this gap by introducing a family of uncertainty relations that bound bipartite entanglement monotones from below via elements of a quantum Fisher information matrix. To further emphasize the significance of our results, we connect these relations to the achievable precision in multiparameter estimation. Considering a system split into two parts with arbitrary dimension, we also show that, while two-dimensional entanglement is sufficient to estimate a single parameter with maximal precision, genuine high-dimensional entanglement is required for multiparameter estimation. We conclude by illustrating how our method extends naturally to a multipartite splitting.
\end{abstract}

\maketitle

\textbf{Introduction.---}The geometry of the quantum state space can be understood in terms of the quantum Fisher information (QFI) metric \cite{PhysRevD.23.357,PhysRevLett.72.3439}, which leads to the Bures distance~\cite{bures1969extension,book,sommers2003bures}. This also has practical advantages, as the QFI is linked to important quantities associated with quantum state space dynamics, such as linear response theory~\cite{HaukeHeylTagliacozzoZoller16} and quantum speed limits~\cite{toth2014quantum,TaddeiQuantum2013,SebastianQuantumJPA2017}. Moreover, the QFI, which can be seen as a generalized and convex variance~\cite{toth2013extremal,yu2013quantum}, is connected to statistical and information-theoretic concepts of quantum theory, such as parameter estimation~\cite{giovannetti2011advances,QuantumPezze2018,NOONZhang2018,QuantumVittorioScience2004,GessnerSensitivity2018}, uncertainty relations~\cite{T_th_2022,WatanabeUncertainty2011}, and quantum resources~\cite{TanKokChuanFisher2021,MarvianOperational2022,TanCoherence2018,YamaguchiBeyond2023}, especially entanglement~\cite{NanEntanglement2013,LucaEntanglement2009,Zhang_2020,HyllusFisher2012,TothMultipartite2012,DetectingHong2015,RenMetrological2021,szalay2024alternativesentanglementdepthmetrological,XingMeasurePRA2019}. In fact, because of its connection with the QFI and the quantum Cram\'er-Rao bound, entanglement turned out to be a crucial resource for quantum metrology, as noted starting from the seminal work~\cite{LucaEntanglement2009}, and subsequent investigations~\cite{HyllusFisher2012,TothMultipartite2012,giovannetti2011advances,toth2014quantum,QuantumPezze2018}. 

This motivated further extensive research in multiparticle-entanglement-enhanced metrology. Recent results include bipartite~\cite{NanEntanglement2013,AoXiangEntanglement2024}, as well as multipartite entanglement criteria~\cite{LucaEntanglement2009,Zhang_2020,ImaiMetrological2025} based on the QFI. These also include witnesses of $k$-producibility~\cite{HyllusFisher2012,TothMultipartite2012,SuDynamical2025} and $k$-separability~\cite{DetectingHong2015}, as well as unifications of these latter concepts~\cite{RenMetrological2021,szalay2024alternativesentanglementdepthmetrological}. Furthermore, the roles of mode and particle entanglement, as well as their interplay, have been investigated in multiparameter and distributed sensing ~\cite{GessnerSensitivity2018,MultiparameterMatteoNJP2023,WangExact2025,HongQuantum2025,zang2025quantumadvantagedistributedsensing} and in quantum walks \cite{CamachoQuantum2025}. 

At the same time, entanglement is often operationally quantified by functions that cannot increase under so-called local operations and classical communication (LOCC), which are known as {\it entanglement monotones}. In bipartite systems, there are many examples of LOCC monotones, often related to entropies of single-party marginals, such as the Schmidt number, the linear entropy of entanglement, the concurrence, the entanglement of formation and others, or related to the violation of the positive partial transpose (PPT) condition~\cite{PlenioVirmani07,HorodeckiEntanglementReview2009}. Many of those entanglement monotones also acquire operational significance in different contexts. For example, the entanglement of formation, as well as the logarithmic negativity, {provides bounds} to the distillable entanglement and in turn a quantification of the usefulness of the state for quantum communication, especially in an asymptotic setting \cite{BennettMixed1996,PlenioLogarithmic2005,leditzky2017useful}. Similarly, the zero-entropy of entanglement, also {known as the Schmidt number or {\it entanglement dimensionality}, has been proven to be a useful benchmark for single-shot tasks} in quantum communication and quantum computation~\cite{erhard2020advances,VandenNestUniversal2013,EckerOvercoming2019,BechmannPasquinucciQuantum2000,CerfSecurity2002,MarioTwistedPNAS2015,AliciaHighOptica2017,CozzolinoOrbitalPRAppl2019}. This is because, roughly speaking, it has the peculiarity of being discrete and giving a more appropriate characterization of the minimal ``memory'' resource needed to simulate quantum states in a computational sense~\cite{VandenNestUniversal2013}. Relations among these entanglement monotones have also been investigated, both operationally~\cite{PhysRevLett.106.130503} {and} conceptually~\cite{HorodeckiEntanglementReview2009,GuheneToth09,FriisNatPhys19,EltschkaSiewert2014}. Given this background, it is natural to ask whether LOCC monotones also have significance in quantum metrology.

In this work, we consider a system composed of two qu$d$its of arbitrary dimension and introduce a method to construct uncertainty relations to bound entanglement monotones based on the quantum Fisher information matrix (QFIM), understood as a convex analogue of the covariance matrix and constructed from local orthogonal bases of the two parties. In a system composed of $N$ particles (e.g., qubits), this translates into an analysis of entanglement across bipartitions. In particular, we find a number of inequalities that bound entanglement monotones via traces of the QFIM of $su(d)$ operators, which are related to the estimation of a full $(d^2-1)$-parameter $SU(d)$ unitary. 

From the same quantities, we can also {\it define} an LOCC monotone, which is then related to the commonly discussed ones in the literature. Our results can then be translated into a direct connection between LOCC monotones and precision bounds in multiparameter estimations with a unitary encoding. In this setting, a further concrete observation arising from our analysis is that, on the one hand, two-dimensional entanglement is sufficient to achieve maximal precision for the estimation of a single parameter. On the other hand, a larger value of LOCC monotones can allow for a lower sum of mean square errors when estimating multiple parameters at once. Finally, we also discuss how to extend our method to the analysis of {multipartite} scenarios, i.e., considering all bipartitions of an $N$-particle system at the same time.%, which connects genuine-multipartite and high-dimensional entanglement to traces of QFIM blocks.

Different from the analysis of multipartite entanglement based on, e.g., the entanglement depth, which roughly quantifies how many particles are entangled with each other in a multiparticle system, in our case we focus on LOCC monotones across bipartitions, which are generalized into a vector of such quantities associated to all the bipartitions. This in turn provides complementary information with respect to the non-$k$-producibility or non-$k$-separability. With our results, we thus complement existing literature in the context of entanglement structure via QFI, as illustrated via an example in the End Matter (see also Supplemental Material~\cite{supplement}).

\textbf{Methods.---}Let us consider a bipartite system with Hilbert space $\hil_a \otimes \hil_b$ of dimension $d_a \times d_b$. A pure bipartite state can be put in a canonical form under local unitaries called{\it Schmidt decomposition}: $\ket{\psi}=\sum_{i=1}^{r} \sqrt{\lambda_i}\ket{u_i}\ket{v_i}$. The corresponding {\it Schmidt coefficients} $\vec \lambda := \{\lambda_i \}_{i=1}^{r}$ form a probability distribution and are used in the definitions of LOCC monotones for pure states~\cite{HorodeckiEntanglementReview2009,GuheneToth09,FriisNatPhys19}. A typical way to extend these definitions to mixed states is via convex-roof~\cite{uhlmann1998entropy,vidal2000entanglement}: 
\begin{equation}\label{eq:convroof}
E(\varrho)=\inf_{\mathcal{D}(\varrho)} \sum_k p_k E\left(\psi_k\right),
\end{equation}
where $E(\psi)$ is defined for pure states and the infimum is over all pure state decompositions $\mathcal{D}(\varrho)=\{p_k,\ket{\psi_k}: \varrho=\sum_k p_k \ket{\psi_k}\bra{\psi_k}\}$. A particularly important example is the entanglement of formation 
$E_F(\psi)=- \tr(\varrho_a \log \varrho_a) = -\sum_{i=1}^r\lambda_i\log \lambda_i$  \cite{BennettMixed1996,HorodeckiEntanglementReview2009}. Other prominent examples include the linear entropy of entanglement 
$E_L(\psi)=1-\tr(\varrho_a^2)$, the 2-tangle $\mathcal{T}\left(\psi\right)=2 E_L\left(\psi\right)$, and the concurrence $C(\psi)=\sqrt{2[1-\tr(\varrho_a^2)]}=\sqrt{2 E_L(\psi)}$ \cite{MintertConcurrence2004,rungta2001universal,rungta2003concurrence}. Alternatively, one can construct entanglement monotones, for example, from the negativity under partial transpose $\mathcal N(\varrho)
=(\|\varrho^{T_b}\|_1 - 1)/2$~\cite{vidal2002computable}. 

One exceptional discrete, rather than continuous, case is given by the Schmidt number
\footnote{For pure states this is usually called Schmidt rank.} 
$\mathcal{SN}(\psi)$, i.e., the number of nonzero Schmidt coefficients. For a mixed state, it is defined as~\cite{SchmidtBarbaraPRA2000},
\begin{equation}
\mathcal{SN}(\varrho):=\inf_{\mathcal{D}(\varrho)} \max_{\left|\psi_k\right\rangle \in \mathcal{D}(\varrho)} \mathcal{SN}(\psi_k) .
\end{equation}

Entanglement monotones can also be related to each other. For example, given any convex and monotonically increasing function $V(x)$ such that $V(C(\psi)) \leq E_F(\psi)$, we have that a bound on the concurrence for pure states $C(\psi) \geq \mathcal B$ can be used to bound the entanglement of formation as~\cite{MintertConcurrence2004}
\begin{equation}\label{eq:eoffromconc}
E_F(\varrho) \geq V(\mathcal B) .
\end{equation}
In particular, the optimal function $V(x)={\rm co} (R_L^{(d)}(x))$ is calculated in Refs.~\cite{LiFeiPRA2010,ObservableZhang2011}, and also given explicitly in the SM~\cite{supplement}. Other relations can also be found, e.g., a bound on the concurrence from one on the negativity~\cite{ChenPRL05,deVicenteLower2007}, or a bound on the Schmidt number from one on the concurrence or the negativity~\cite{PhysRevA.91.032327}.

Next, let us introduce the quantum Fisher information matrix (QFIM)~\cite{JingQuantum2019}. Consider a family of quantum states generated via a unitary encoding $\varrho_{\vec \theta} = U_{\vec \theta} \varrho_0 U^\dagger_{\vec \theta}$, where $U_{\vec{\theta}}=\exp \left(-\mathrm{i} H(\vec\theta) \right)$ and $\vec \theta$ denotes the vector of parameters.
The QFIM can be defined as 
\begin{equation}
\label{eq:QFIMunitary}
[\mathcal F_{\varrho_{\vec \theta}}]_{jk} =\sum_{l,m,\lambda_l+\lambda_m\ne 0}\frac{2(\lambda_l-\lambda_m)^2}{\lambda_l+\lambda_m}\Re{\bra{l} \tilde H_j \ket{m} \bra{m} \tilde H_k \ket{l}},
\end{equation}
where $\varrho_0= \sum_l \lambda_l \ketbra{l}$ is the spectral decomposition of the density matrix and the generators $\tilde{\vec{H}}$ are obtained from the unitary evolution as $\tilde H_k = {\rm i} (\partial_k U_{\vec{\theta}}^\dagger) U_{\vec{\theta}} = - \int_0^1 e^{{\rm i}s H(\vec\theta)} (\partial_k H(\vec\theta)) e^{-{\rm i}s H(\vec\theta)} \de s$~\cite{JingQuantum2019}. By implicitly referring to such a unitary parametrization, one can then define the QFIM $\mathcal F_{\varrho}(\vec M)$ relative to a quantum state $\varrho$ and an arbitrary vector of observables $\vec M \equiv \tilde{\vec{H}}$, which is what we do in the following. Note also that the QFI can be seen as a generalized variance, which for pure states is proportional to the variance and for mixed states is obtained via convex-roof~\cite{toth2013extremal,yu2013quantum}. 

In the following, we focus on local operators given by full orthonormal bases $\vec g_a$ and $\vec g_b$ for the two parties. A typical choice is given by the normalized identity $g_1^{(n)} = \id/\sqrt d$ plus an $su(d)$ basis normalized as $\tr(g^{(n)}_k g^{(n)}_l)=\delta_{kl}$, where $n=a,b$ and we consider $d_a=d_b=d$. For a bipartite density matrix $\varrho$, the QFIM relative to the vector $\vec g = (\vec g^{(a)}, \vec g^{(b)}):=(\{g^{(a)}_k \otimes \id\}_{k=1}^{d^2},\{\id \otimes g^{(b)}_k\}_{k=1}^{d^2})$ has the form: 
\begin{equation}\label{eq:QFIMBlockDef}
\mathcal{F}_\varrho(\vec g)=\left(\begin{array}{cc}
\mathcal{F}_a & \mathcal X_\varrho \\
\mathcal X^T_\varrho & \mathcal F_b
\end{array}\right),
\end{equation}
where $\mathcal{F}_a := \mathcal{F}_{\varrho}(\vec g^{(a)})$ and $\mathcal F_b := \mathcal{F}_{\varrho}(\vec g^{(b)})$ are QFIMs of local operators, while $\mathcal X_\varrho$ represents the cross block between the two parties. Note also the analogy with the covariance matrix $[\Gamma_{\varrho}(\vec g)]_{jk} = \tfrac 1 2 \av{g_j g_k + g_k g_j}_\varrho - \av{g_j}_\varrho \av{g_k}_\varrho$, which can be used for entanglement~\cite{guhnecova,gittsovich08,GittsovichPRA10} as well as Schmidt number detection~\cite{Liu2024bounding,liu2022characterizing,NonlinearLiuArxiv2024}. 

Let us recall that the QFIM is a real, symmetric and positive semi-definite matrix that is convex under mixing the quantum state, i.e., $\mathcal{F}_{p \varrho_1+(1-p) \varrho_2}(\vec g) \preceq p \mathcal{F}_{\varrho_1}(\vec g) +(1-p) \mathcal{F}_{\varrho_2}(\vec g)$~\cite{JingQuantum2019}. For pure states, the QFIM becomes four times the covariance matrix, i.e., $\mathcal{F}_{\psi}(\vec g) = 4\Gamma_\psi(\vec g)$. 

\textbf{Results.---} Let us now try to characterize entanglement monotones via the QFIM. The key idea is to exploit its convexity, {together with the known characterization of pure-state covariance matrices} to bound convex-roof entanglement monotones. From \cref{eq:convroof}, we can see that the {values of these entanglement monotones are obtained} from an optimal decomposition into pure states $\varrho = \sum_k p_k^{(o)} \ketbra{\psi_k^{(o)}}$. One can then use the convexity of QFIM $\mathcal{F}_\varrho(\vec g) \preceq \sum_k p_k \mathcal{F}_{\psi_k}(\vec g)=4 \sum_k p_k \Gamma_{\psi_k}(\vec g)$ and try to characterize the boundary covariance matrix $\Gamma_{\psi_k}(\vec g)$ for $\ket{\psi_k}$ with a given value of some entanglement monotone $E(\psi_k)$. This can be done due to the fact that the covariance matrix of a pure state can be conveniently expressed in an operator basis coming from the Schmidt bases~\cite{GittsovichPRA10,Liu2024bounding}. 

One can take a scalar combination of the QFIM elements and derive the corresponding bound coming from $\Gamma_{\psi_k}$. In particular, we consider the quantity
\be\label{eq:Tfunctiondef}
T^{(t)}(\varrho)=\tr\left(\mathcal{F}_a\right)+t^2\tr\left(\mathcal{F}_b\right)+2 t\cdot\tr\left|\mathcal{X}_\varrho\right| ,
\ee
where $\tr\left|\mathcal{X}_\varrho\right| := \|\mathcal{X}_\varrho\|$ is the trace norm, which is also the norm that we use throughout the work. \cref{eq:Tfunctiondef}
arises from the trace norm of the blocks of $\mathcal{F}_\varrho(\vec g)$. Here $t\geq 0$ and we denote $T(\varrho)\equiv T^{(1)}(\varrho)$.
For pure states, this quantity is a linear combination of entanglement monotones, and thus its convex roof 
\be
E^{(t)}_T(\varrho) = \inf_{\mathcal{D}(\varrho)} \frac 1 8 \sum_k p_k T^{(t)}(\psi_k) -\frac{t^2+1}{2}(d-1) ,
\ee
is also a valid entanglement monotone for mixed states (See \smcref{sec:ContinuousMonotoneProof}). 
Here, the subtracted term makes it vanish for separable states. For the special cases $t=0$ or $t\rightarrow \infty$, such a monotone reduces to the $2$-tangle~\cite{supplement}.

Due to the above convex-roof definition and the convexity of $T^{(t)}(\varrho)$ which is inherited from the QFIM, we also find the bound
\begin{equation}\label{eq:ETtBound}
E^{(t)}_T(\varrho) \geq \frac18T^{(t)}(\varrho) -\frac{t^2+1}{2}(d-1).
\end{equation}

The function $T(\varrho)$ can also be bounded from below by a sum of collective QFIs:
\be\label{eq:TboundsumFGi}
T(\varrho) \geq \sum_{i=2}^{d^2} F_\varrho(G_i) ,
\ee
where $G_i=g_i^{(a)} \otimes \mathbb{1} + \mathbb{1} \otimes g_i^{(b)}$ are collective basis operators and generically the identity will not contribute, thus we can restrict the sum to $(d^2-1)$ $su(d)$ operators (See \smcref{sec:ContinuousMonotoneProof}). Based on this quantity we can derive bounds on several convex-roof entanglement monotones with the following general idea.

Consider a bipartite state $\varrho$ and any convex-roof entanglement monotone $E(\varrho)$. For every pure-state decomposition $\left\{p_k,\left|\psi_k\right\rangle\right\}$ of $\varrho$, one has $E(\varrho)\leq \sum_k p_k E(\psi_k)$, and there exists an optimal decomposition $\{p_k^{(o)},|\psi_k^{(o)}\rangle\}$ that saturates this bound. Therefore, any lower bound on $\sum_k p_k E\left(\psi_k\right)$ valid for all pure-state decompositions immediately gives a corresponding lower bound on $E(\varrho)$.

The general derivation is, in principle, applicable for any such entanglement monotone. Here, we showcase representative examples, starting with the concurrence and, consequently, the entanglement of formation, which also has a clearer operational significance.

\begin{observation}\label{observation2}
For a bipartite system of dimension $d\times d$ the following bounds hold 
\begin{subequations}\label{eq:CVMonotones}
\begin{align}
C(\varrho) &\geq \frac{\sqrt{d}(T(\varrho)-8(d-1))}{4(d+2) \sqrt{2(d-1)}}:=\mathcal{B}_C(\varrho), \label{eq:ConcurrenceBound}\\
E_{F}(\varrho) &\geq V\left(\mathcal{B}_C(\varrho)\right), \label{eq:EoFBound}
\end{align}
\end{subequations}
where $T(\varrho)$ is given by \cref{eq:Tfunctiondef} and the convex function $V(x)$ mentioned in \cref{eq:eoffromconc} is given explicitly in the SM~\cite{supplement}.
\end{observation}

\textbf{Proof.---} 
For pure states, we have $\tr\left|\mathcal{X}_\psi\right|=E_L(\psi)+2 \mathcal{N}(\psi)$ and $\tr\left(\mathcal{F}_a\right)=\tr\left(\mathcal{F}_b\right)=4\left[(d-1)+E_L(\psi)\right]$. Using $[C(\psi)]^2=2 E_L(\psi)$ and the Cauchy-Schwarz inequality $\mathcal{N}(\psi) \leq \sqrt{\frac{d(d-1)}{2}} \frac{C(\psi)}{2}$ and noting $C(\psi)\le \sqrt{\frac{2(d-1)}{d}}$, we recover \cref{eq:ConcurrenceBound}. Then, we use the relation $E_F\left(\psi_k\right) \geq \operatorname{co}\left(R_L^{(d)}\left(C\left(\psi_k\right)\right)\right)$ from~\cite{ObservableZhang2011,DetectingZhang2013} to further bound the entanglement of formation. Since this relation holds for all pure states, we can extend it to the optimal decomposition of $\varrho$ due to the convexity of the bound. See~\smcref{sec:ContinuousMonotoneProof} for details.
\qed

For the discrete entanglement dimensionality, even though the exact same method does not apply, we can still use similar techniques and obtain:
\begin{observation}\label{observation1}
For a bipartite state $\varrho$ of dimension $d\times d$ with a Schmidt number $\mathcal{SN}(\varrho)$ not exceeding $r$, the following inequalities must hold
\begin{subequations}\label{eq:5}
\begin{gather}
  \max \{ \tr(\mathcal F_a) , \tr(\mathcal F_b) \}   \leq 4 \left( d - \tfrac 1 r \right), \label{eq:addconds1} \\
      \frac{\tr|\mathcal X_\varrho|} 4 - \sqrt{\left( d-\frac{1}{r}-\frac{\tr(\mathcal F_a)} 4\right)\left(d-\frac{1}{r}- \frac{\tr(\mathcal F_b)} 4\right)} \leq r - \frac 1 r . 
        \label{eq:FXTraceNormUpperBound}
    \end{gather}
\end{subequations}
A violation of any of them, in particular \cref{eq:FXTraceNormUpperBound}, implies that the quantum state must have a Schmidt number larger than $r$. 
\end{observation}

$T^{(t)}(\varrho)$ and the criteria in \cref{eq:5} are independent of the choice of the local basis, because changes in the local basis correspond to local orthogonal transformations of the form $(O_a \oplus O_b) \mathcal{F}_\varrho(\vec g) (O_a \oplus O_b)^T$, which leave the expressions invariant. It is noteworthy that the bound in \cref{eq:FXTraceNormUpperBound} is not only saturated by Schmidt-rank-$r$ maximally entangled pure states, but also by certain mixed states (See \smcref{sec:ObsAlternativeProof}).

Remarkably, \cref{eq:addconds1} is based solely on the QFIMs of local operators, even though they are calculated from the global bipartite state. The condition in \cref{eq:addconds1} follows from the bound on the local purities of each pure state component $\ket{\psi_k}$.
A similar expression also provides a bound on the $2$-tangle as follows:

\begin{observation}\label{observation3}
For a bipartite system of dimension $d\times d$ with $2$-tangle $\mathcal{T}(\varrho)$, the following inequality holds
\begin{equation}
\label{eq:LinearEntropyBound}
2\mathcal T(\varrho) \geq \max\{ \tr(\mathcal F_a), \tr(\mathcal F_b)\}-4(d-1) .
\end{equation}
\end{observation}

An extended proof of this statement can be found in the SM~\cite{supplement}. Here we just
note that \cref{eq:LinearEntropyBound} can also be obtained from \cref{eq:ETtBound} by setting $t=0$ and $t\rightarrow \infty$.

In the following summarizing statement, we readily obtain from \cref{eq:TboundsumFGi} uncertainty-relation criteria which are simpler and more intuitive than \cref{eq:ConcurrenceBound,eq:EoFBound,eq:FXTraceNormUpperBound,eq:ETtBound}, i.e., essentially by substituting the trace norm with the trace. At the same time, the right-hand side of \cref{eq:FXTraceNormUpperBound} can be further upper-bounded by using the relation $\sqrt{xy} \leq \tfrac 1 2 (x+y)$.

\begin{corollary}\label{corollary1}
The following bounds hold
\begin{equation}\label{eq:FisherTraceSum}
\sum_{i=2}^{d^2} F_\varrho(G_i) \leq \left\{ 
\begin{array}{l}
8\left( d +\mathcal{SN}(\varrho)-\frac{2}{\mathcal{SN}(\varrho)}\right) \\
\frac{4(d+2) \sqrt{2(d-1)}}{\sqrt{d}} C(\varrho)+8(d-1) \\
\frac{4(d+2) \sqrt{2(d-1)}}{\sqrt{d}} V^{-1} \left(E_F(\varrho)\right)+8(d-1) \\
8E_T^{(1)}(\varrho)+8(d-1)
\end{array}
\right. \,,
\end{equation}
where $G_i=g_i^{(a)} \otimes \mathbb{1} + \mathbb{1} \otimes g_i^{(b)}$ with $\{g_i^{(a/b)}\}_{i=1}^{d^2}$ forming local bases for the two parties and $V^{-1}$ is the inverse function of that mentioned in \cref{eq:eoffromconc}.
\end{corollary}

\begin{figure}[h]
\centering
\includegraphics[width=\columnwidth]{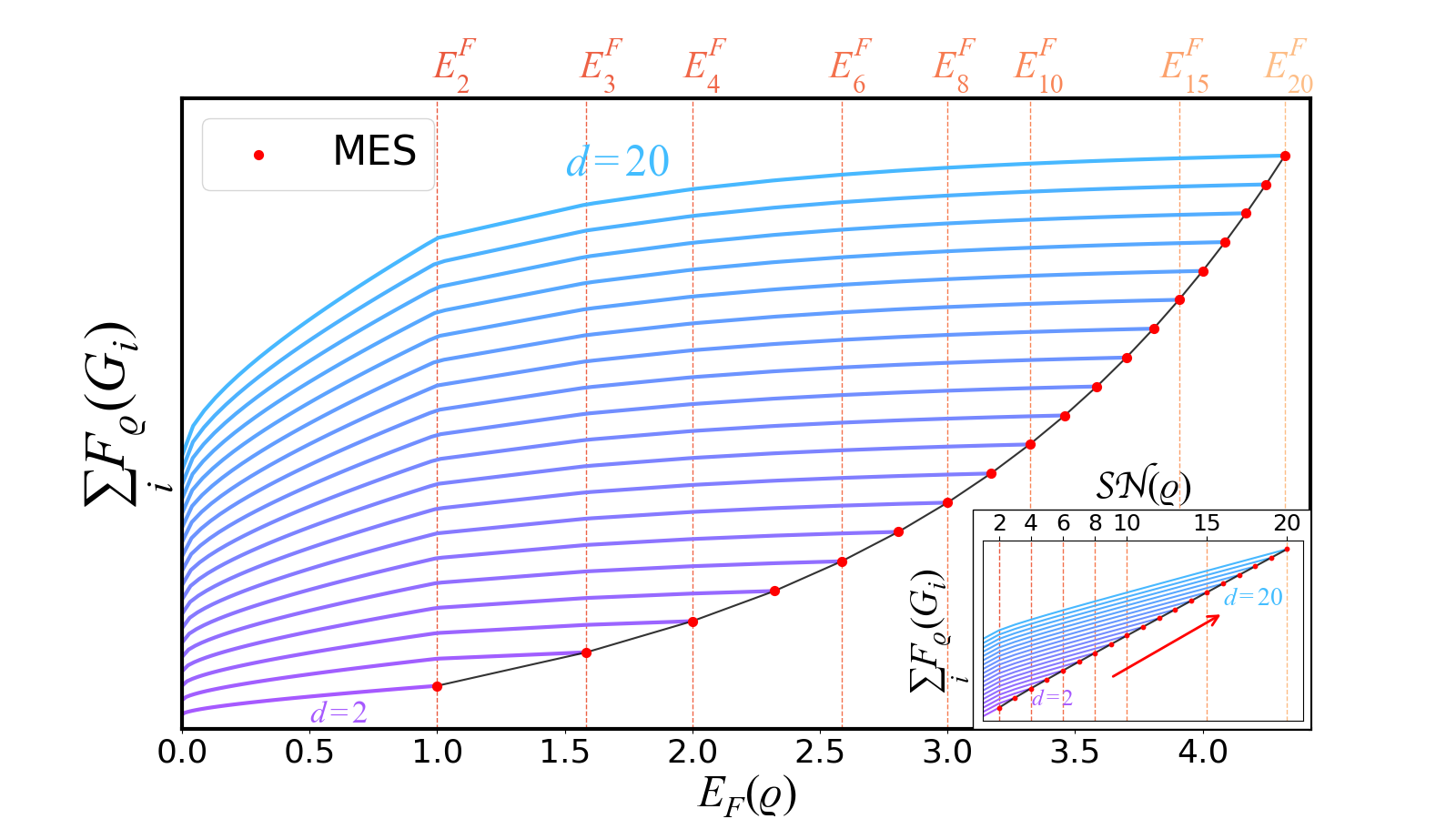}
\caption{Bounds on the sum of collective QFIs, $\sum_i F_\varrho(G_i)$, as functions of two typical entanglement monotones: the entanglement of formation $E_F(\varrho)$ and the Schmidt number $\mathcal{SN}(\varrho)$. In the main and inset plots, we consider all bipartite high dimensional states from $d=2$ to $d=20$. The blue-purple gradient curves represent the maximal value of $\sum_i F_\varrho(G_i)$ for a fixed monotone value. Red points on the solid black curves correspond to maximally entangled states (MES) of different dimensions. The vertical dashed lines $E_d^F$ denote the maximum monotone values of $d$-dimensional states as a reference. The inset plot of Schmidt number is drawn likewise.
}
\label{fig:FQAndDiscreteContinuousMonotones}
\end{figure}

See \cref{fig:FQAndDiscreteContinuousMonotones} for a plot of the bounds above. 

From their derivation, we also see that \cref{eq:FisherTraceSum} is weaker than \cref{eq:FXTraceNormUpperBound,eq:CVMonotones}. Moreover, the usefulness of \cref{eq:FisherTraceSum} depends on the choice of the bases $(\vec g_a, \vec g_b)$. However, when $\tr(\mathcal F_a)=\tr(\mathcal F_b)$ and for an optimal choice of bases, \cref{eq:FXTraceNormUpperBound,eq:CVMonotones} become equivalent to \eqref{eq:FisherTraceSum}. Such an optimal pair of local bases would correspond to the one that gives the singular value decomposition of $\mathcal X_\varrho$ (see also the proof of \cref{observation1} in \smcref{sec:ObsAlternativeProof} for more details).

We also note that a similar criterion was derived in Ref.~\cite{NanEntanglement2013,AltenburgSchemes2017Thesis} for deciding if a state is entangled or not. Thus, our results, in particular \cref{corollary1}, can be seen as a direct generalization of those works to quantify entanglement via LOCC monotones.

These uncertainty relations are based on the sum of $(d^2-1)$ collective QFIs. Next, it is natural to ask whether one can derive similar criteria based on a single QFI corresponding to a collective observable of the form $M=A\otimes \id + \id \otimes B$, as it is the case for just entanglement (e.g., distinguishing $r=2$ and $r>2$ in terms of the Schmidt number). In fact, in some cases the relation between QFI and variance, like $F_\varrho(M)\leq 4 (\Delta M)^2_\varrho$ can be exploited to derive uncertainty relations~\cite{Chiew_2022,T_th_2022} and entanglement criteria~ \cite{LucaEntanglement2009,HyllusFisher2012,TothMultipartite2012,GessnerPezzeSmerzi16,Zhang_2020,MultiparameterMatteoNJP2023}, even with a single observable $M$. Such a collective observable can be seen as a vector $\vec m = (\vec a, \vec b)$ when expressed in the bases $(\vec g_a,\vec g_b)$ and the corresponding QFI is obtained as $F_\varrho(M)= \vec m \cdot \mathcal{F}_\varrho(\vec g) \cdot \vec m^T$. Then, one can look at the maximum of $F_{\psi_r}(M)$, e.g., over all pure states with Schmidt rank given by $r$. However, we find that such a bound is trivial already in the case of qubit-level entanglement. See End Matter for a more precise statement and SM for further details.

Here, we provide an illustrative example that shows how multiple QFIs can reveal high-dimensional entanglement that is not visible with a single QFI. Let us consider the following $4$-qubit state:
\begin{equation}
|\Phi\rangle=\frac{\rm i}{\sqrt{2}}\left|\psi^{-}\right\rangle_{12}\left|\psi^{-}\right\rangle_{34}+\frac{1}{\sqrt{6}}\left|\psi^{-}\right\rangle_{13}\left|\psi^{-}\right\rangle_{24}+\frac{1}{\sqrt{6}}\left|\psi^{-}\right\rangle_{14}\left|\psi^{-}\right\rangle_{23} ,
\end{equation}
where $\left|\psi^{-}\right\rangle=\frac{1}{\sqrt{2}}(|01\rangle-|10\rangle)$. Being a superposition of products of two-particle singlet states, we have $F_{\Phi}(J_k)=4(\Delta J_k)^2_{\Phi}=0$, where $J_k=\tfrac12\sum_{\ell=1}^4 \sigma_k^{(\ell)}$ are total spin operators with $k=x,y,z$. Consequently, any single-$J_k$-based criterion, including the $k$-producible upper bounds of~\cite{HyllusFisher2012,TothMultipartite2012}, returns a trivial result ($F_Q=0$) and cannot detect any entanglement in $\ket{\Phi}$. We note that tailoring the local operators independently for each particle, a single global QFI witness {\it can} detect entanglement in the state, but only at the qubit level. See SM, below \cref{eq:PhiBellFactorize}.

Nevertheless, $\ket{\Phi}$ is genuinely $4$-partite entangled (entangled across every bipartition, e.g. composed of $1$ vs $3$ particles) and exhibits Schmidt rank $4$ across all $2$ vs $2$ bipartitions. In fact, our QFIM-based criterion also yields $\mathcal{SN}(\ket{\Phi})=4$. Additionally, our EoF criterion in \cref{eq:EoFBound} gives $E_F(\ket{\Phi}) \geq 1.4025$, which, in particular, exceeds the Bell-state value ($1$ ebit).

Our observations can also be generalized to quantify genuine multipartite entanglement, e.g., via a vector $\vec{\mathcal{E}}(\varrho)$ whose entries are the entanglement entropies for every bipartition~\cite{HuberStructure2013,malik2016multi,erhard2018experimental,CerveraExperimental2022,hu2020experimental,bao2023very,klockl2015characterizing,NonlinearLiuArxiv2024}. In the End Matter and in \smcref{sec:MultipartiteEntanglement} we showcase how to extend the Schmidt number criterion given in \cref{eq:FXTraceNormUpperBound} from the bipartite case to the multipartite scenario. The current literature focuses on bounds on $k$-producibility (entanglement depth) and $k$-separability \cite{RenMetrological2021}. The same information can also be extracted from the entanglement dimensionality (Schmidt number) vector. See End Matter and \smcref{sec:MultipartiteEntanglement} for an example of how our method detects non-$k$-producibility and non-$k$-separability of pure states.

{\bf Connections to QCRB.---}Finally, let us observe some implications that our results have on the relationship between entanglement monotones and phase-estimations. In fact, exploiting the usual quantum Cramér-Rao bound (QCRB), i.e., the fact that the precision of estimation is lower-bounded by the inverse of the QFI, it is possible to lower-bound the QFIs from the sensitivity of the corresponding estimation task obtained, e.g., via error-propagation formula as a signal-to-noise ratio of a measured observable. More precisely, the QFI of an observable $H$ can be estimated by referring to the unitary parameter encoding $U(\theta) = \exp(-i\theta H)$ and, for example, considering the estimation via the measurement of a $\theta$-sensitive observable $O(\theta)$ and $\theta \simeq 0$ as $(\delta \theta)^{-2} = \frac{|\av{[O,H]}_\varrho|^2}{(\Delta O)_\varrho^2} \leq F_\varrho(H)$, where the sensitivity $(\delta \theta)^{-2} = \frac{|\av{[O,H]}_\varrho|^2}{(\Delta O)_\varrho^2}$ is obtained via the usual error-propagation formula~\cite{toth2014quantum,QuantumPezze2018}. Thus our method, in particular inequalities like \eqref{eq:FisherTraceSum}, can be evaluated by further bounding the left-hand side via signal-to-noise ratios of suitable observables, as this relation can be also extended to the multiparameter case~\cite{GessnerSensitivity2018,gessnerNatComm20}.

At the same time, the QCRB, together with our results, i.e., criteria like \cref{eq:FisherTraceSum}, also implies a bound to the sensitivity of multiparameter estimation using quantum states with bounded entanglement, as quantified via LOCC monotones. Let us illustrate this relation considering a multiparameter unitary encoding of the form {$U(\vec \theta)=e^{-{\rm i}\sum_k \theta_k G_k}$}, where again $G_k =g_k^{(a)}\otimes \id + \id \otimes g_k^{(b)}$, and considering as a figure of merit the total estimation variance given by $\sum_{i=2}^{d^2} (\delta \theta_i)^2$, {calculated near the point $\vec \theta = (0,\dots ,0)$}. Combining criteria \cref{eq:FisherTraceSum} with the bound 
\be
\sum_{i=2}^{d^2} (\delta \theta_i)^2 \geq \frac{(d^2-1)^2}{\sum_{i=2}^{d^2} F_{\varrho}(G_i)} ,
\ee
we thus obtain that, for every entanglement monotone considered, greater entanglement lowers the total mean square error in multiparameter estimation tasks. See \smcref{sec:Lemma1Proof} for a proof of this inequality.

{\bf Conclusions.---}In conclusion, we introduce a family of uncertainty relations based on QFIM and LOCC entanglement monotones, which generally show how the precision limit of multiparameter estimation increases as entanglement increases. We also propose a general approach for extending our observations to multipartite systems, enabling the detection of the entanglement-entropy vector. Our work raises several interesting questions that deserve further investigation, including, for example, the relation between the QFIM and the covariance matrix criteria and to other entropic entanglement measures that do not rely on convex-roof constructions. Another interesting problem would be to systematically find inequalities of the form \eqref{eq:FisherTraceSum} or \eqref{eq:boundsforfewQFIs}, especially for the multipartite case, eventually optimized for specific situations like numerical investigation of many-body models or implementation in experiments \cite{EntanglementQuantificationFadel2021}. It is worth pointing out that in the particular case of thermal equilibrium states, our criteria can be also evaluated from linear response functions, due to the relation between the QFI and dynamical susceptibilities~\cite{HaukeHeylTagliacozzoZoller16}.

We thank Zhen-Peng Xu, G\'eza T\'oth, Otfried G\"uhne, Satoya Imai, Manuel Gessner and Augusto Smerzi for discussions. This work is supported by Quantum Science and Technology-National Science and Technology Major Project (Grants No. 2024ZD0302401 and No. 2021ZD0301500), National Natural Science Foundation of China (No. 12125402, No. 12534016, and No. 12405005), and Beijing Natural Science Foundation (Grant No. Z240007). S.L. acknowledges the China Postdoctoral Science Foundation (No. 2023M740119). M.F. was supported by the Swiss National Science Foundation Ambizione Grant No. 208886, and by The Branco Weiss Fellowship--Society in Science, administered by the ETH Z\"{u}rich. G.V. acknowledges financial support from the Austrian Science Fund (FWF) through Grants P 35810-N and P 36633-N (Stand-Alone).

$^{\ddagger}$S. Du and S. Liu contributed equally to this work.

{\bf End Matter.---}\textbf{Proof. of \cref{observation1}---}
Assuming a bipartite state $\varrho$ has Schmidt number $r$, there must exist a pure-state decomposition $\varrho=\sum_k p_k |\psi_k \rangle\langle \psi_k |$ in which each $\left|\psi_k\right\rangle$ has Schmidt rank at most $r$. This, together with convexity of the QFIM, leads to the matrix inequality $\mathcal{F}_\varrho(\vec g) \preceq \sum_k p_k \mathcal{F}_{\psi_k}(\vec g)=4 \sum_k p_k \Gamma_{\psi_k}(\vec g):=4\Gamma_r$. The general form of the boundary matrix $\Gamma_r:=\sum_k p_k \Gamma_{\psi_k}(\boldsymbol{g})$ is also known~\cite{gittsovich08,GittsovichPRA10,Liu2024bounding}. Positivity of the matrix $\Delta := 4\Gamma_r - \mathcal{F}_\varrho(\vec g)$ implies that $\tr(|\Delta_a|) \tr(|\Delta_b|) \geq (\tr|\Delta_X|)^2$ where we indicate by $\Delta_{a/b/X}$ the corresponding blocks. Furthermore, $\Delta_a$ and $\Delta_b$ are principal minors of $\Delta$ and thus they have to be positive semidefinite. Next, we use the bounds $0\leq \tr(\Delta_{a/b}) \leq 4(d-1/r) -\tr(\mathcal F_{a/b})$, which imply \cref{eq:addconds1}. Then, we use the bound $\tr|\Delta_X| = \tr|4X_r - \mathcal X_\varrho| \geq \left|4\tr|X_r| - \tr|\mathcal X_\varrho|\right| \geq \left(\mathcal \tr| \mathcal X_\varrho|- 4\tr|X_r| \right)  \geq \tr|\mathcal X_\varrho | - 4(r-1/r)$, which holds for all $\Gamma_r$ coming from mixtures of Schmidt-rank-$r$ pure states. Substituting all those bounds we get \cref{eq:FXTraceNormUpperBound}. See \smcref{sec:ObsAlternativeProof} for further details and an alternative proof.
\qed

{\bf Schmidt-number bound with fewer QFIs.---}
Consider an arbitrary bipartite state $\varrho$ with any local dimensions, and let $A,B$ be arbitrary local operators and $M=A \otimes \mathbb{1}+\mathbb 1 \otimes B$. Denoting by $\mu_{\max}^a,\mu_{\max}^b$ and $\mu_{\min}^a,\mu_{\min}^b$ their largest and smallest eigenvalues respectively, we have the following upper bounds
\begin{equation}\label{eq:nogo}
F_{\varrho}(M)\le 
\begin{cases}
(\mu_{\max}^a-\mu_{\min}^a)^2+(\mu_{\max}^b-\mu_{\min}^b)^2, & \text{separable}\\ (\mu_{\max}^a+\mu_{\max}^b-\mu_{\min}^a-\mu_{\min}^b)^2, & \text{entangled}
\end{cases} ,
\end{equation}
where $F_{\varrho}(M)$ denotes the QFI of $M$ calculated on $\varrho$. For entangled states, including those that are genuinely high-dimensional entangled, the bound is always saturated by Bell-like states, which only have qubit-level entanglement. See~\smcref{sec:NoGoTheoremProof} for a detailed proof.

As a consequence of \cref{eq:nogo}, we have that genuinely high-dimensional entanglement cannot be certified from a single collective QFI, making it necessary to consider more than one collective observable. In general, one can consider a set of collective observables $\vec M = \{M_k\}_{k=1}^K = \{A_k \otimes \id + \id \otimes B_k\}_{k=1}^K $ and look for upper bounds valid for all quantum states with given values of some entanglement monotone. For example, considering states with $\mathcal{SN}(\varrho) \leq r$, we look for:
\be\label{eq:boundsforfewQFIs}
\sum_k F_{\varrho_r}(M_k) \leq   
4 \max_{\mathcal{SN}(\psi_r)=r} \sum_k (\Delta M_k)_{\psi_r}^2 := \mathcal B_r(\vec M),
\ee
where the maximum is taken over all Schmidt-rank-$r$ pure states. In \smcref{sec:CollectiveSpin}, we present illustrative examples for the case of collective spin operators $J_k^\pm=j_k\otimes \mathbb{1}\pm \mathbb{1}\otimes j_k, k=x,y,z$.

{\bf Generalization to multipartite scenario.---}
In the case of multipartite pure states, the so-called entanglement entropy vector is defined as $\{\mathcal{E}_j^{\downarrow}(\psi)\}_j$, where the index $j$ runs over all bipartitions and $\mathcal{E}_j^\downarrow(\psi)$ denotes the $j$-th largest entanglement entropy of $\ket{\psi}$ across all bipartitions. For mixed states, different convex extensions are possible. Throughout this work, we adopt the definition in terms of an element-wise infimum over pure-state decompositions \cite{HuberStructure2013,NonlinearLiuArxiv2024}, $\mathcal{E}_j^\downarrow(\varrho):=\inf _{\mathcal{D}(\varrho)} \sum_k p_k \mathcal{E}_j^\downarrow(\psi_k)$, where $\mathcal{D}(\varrho)$ denotes all pure-state decompositions of $\varrho$. Below, we showcase how zero-entropy (i.e., the Schmidt number) can be used to extend \cref{eq:FXTraceNormUpperBound} from the bipartite to the multipartite case. A similar derivation can, in principle, be adapted to other convex extensions as well.

In the following, we obtain a bound on the entanglement-dimensionality vector $\mathcal{SN}^\downarrow_j(\varrho)=\inf_{\mathcal{D}(\varrho)} \max_{\ket{\psi_k} \in \mathcal{D}(\varrho)} \mathcal{SN}^\downarrow_j(\psi_k)$ where the notation is as defined above, and show that this bound can be computed via a linear program:

\begin{observation}\label{observationMultipartite}
For an $n$-partite state $\varrho$, consider a candidate non-increasingly ordered vector $\vec v=(r_1 , r_2 , \dots , r_{\mathcal P})$. We call $\vec v$ feasible for $\varrho$ if there exist a pure-state decomposition $\{p_k,\left|\psi_k\right\rangle\}$ and integer vectors $\vec r_k = ((r_1)_k,(r_2)_k, \dots ,(r_{\mathcal P})_k)$, together with permutation matrices $M_k$, such that $\vec r_k \leq M_k \vec v$ element-wise for all $k$ and
    \be\label{eq:obsMultipartite}
     \left\{ \begin{array}{cc}
            h_{1}(\varrho) &\leq \sum_k p_k\left((r_{1})_k-\frac{2}{(r_{1})_k}\right) , \\   
            h_2(\varrho) &\leq \sum_k p_k\left((r_{2})_k-\frac{2}{(r_{2})_k}\right) , \\   
             \vdots \\ 
            h_{\mathcal P}(\varrho) &\leq \sum_k p_k\left((r_{\mathcal P})_k-\frac{2}{(r_{\mathcal P})_k}\right),  
    \end{array}
    \right.
    \ee
where the functions $h_j(\varrho)$ are defined in \cref{eq:SNVCrit} for the bipartition $(j|\bar j)$. Finally, taking the element-wise minimum over all feasible vectors, the resulting vector $\vec v^{\min}=(v_1^{\min},\dots,v_{\mathcal P}^{\min})$ is the detected entanglement-dimensionality vector.
\end{observation}

\textbf{Proof.---}Consider a quantum state $\varrho$ composed of $n$ particles of dimension $d$ and the corresponding QFIM for each bipartition $(j | \bar{j})$, labelled by an integer index 
$j \in \{1,2\dots , \mathcal P\}$ with $\mathcal P = 2^{n-1} -1$ being the total number of bipartitions. This QFIM reads 
\begin{equation}
\mathcal{F}_\varrho(\vec g_j)=\left(
\begin{array}{cc}
\mathcal F_j & \mathcal X_\varrho^{(j|\bar{j})} \\
(\mathcal X_\varrho^{(j|\bar{j})})^T &\mathcal  F_{\bar{j}} 
\end{array}
\right) ,
\end{equation}
where $\vec g_j$ is a basis pair for the partition $(j|\bar j)$. 

Let us now consider a quantum state $\varrho=\sum_k p_k \ket{\psi_k}\bra{\psi_k}$. From the covariance matrix for pure states, combined with the convexity of the QFIM and the fact that it is upper bounded by the covariance matrix, we obtain (cf.~SM \cite{supplement}):
\begin{equation}
\begin{split}
    \label{eq:SNVCrit}
h_{j}(\varrho):= &\frac{1}{4}\tr|\mathcal X_\varrho^{(j|\bar{j})} | -\sqrt{\left(d-\frac{1}{4} \tr(\mathcal F_{j})\right)\left(d-\frac{1}{4} \tr(\mathcal F_{\bar j})\right)} \\ 
& \leqslant \sum_k p_k\left((r_j)_k-\frac{2}{(r_j)_k}\right),
\end{split}
\end{equation}
where $\vec r_k$ is a vector such that each $(r_j)_k$ gives an upper-bound to the Schmidt rank of the pure state $\ket{\psi_k}$ across bipartition $j$.

A state $\varrho$ that, for a given vector $\vec v$, satisfies the system of inequalities \eqref{eq:obsMultipartite} cannot rule out the existence of a pure-state decomposition $\{p_k,\ket{\psi_k}\}$ such that each pure state $\ket{\psi_k}$ in the decomposition has a Schmidt-rank vector $\vec r_k \leq M \vec v$ for some permutation matrix $M$. We therefore call $\vec v$ feasible for $\varrho$. Consequently, \eqref{eq:obsMultipartite} is used to solve a linear feasibility problem and to obtain the set of feasible vectors. Taking the element-wise minimum over this set then yields a rigorous lower bound $\vec v^{\min}=(v_1^{\min},\dots,v_{\mathcal P}^{\min})$ on the elements of the entanglement-dimensionality vector of $\varrho$.
\qed

As an example to clarify this result, one can see that, in a tripartite system with equal dimension $d$, the bound from \cref{observationMultipartite} can be saturated by states with the extremal entanglement-dimensionality vector $\vec{v}=(d,d,d)$. A typical example of such states is the GHZ state $\ket{\Psi_{\text{GHZ}}^d}=\sum_{i=1}^d \frac{1}{\sqrt{d}}\ket{i i i}$. Let us also illustrate how our method complements current literature on the topic with a simple example in the following. As we mentioned, current literature focuses on bounds on the depth of entanglement (i.e., non-$k$-producibility) and the degree of genuine multipartiteness (i.e., non-$k$-separability) \cite{HyllusFisher2012,TothMultipartite2012,SuDynamical2025,DetectingHong2015,RenMetrological2021,szalay2024alternativesentanglementdepthmetrological}. Once again, these two provide complementary information about the entanglement structure of the state.
For pure states, the entire entanglement structure, including both pieces of information, can often be extracted directly from the entanglement-dimensionality vector, as we illustrate here via an example.

Consider the following $7$-qubit state (similar states are also considered in~\cite{RenMetrological2021}):
\begin{equation}\label{eq:7qubitstateEM}
\ket{\phi}=\frac{1}{2\sqrt{2}}(\ket{0}+\ket{1})\otimes(\ket{00}+\ket{11})\otimes(\ket{0000}+\ket{1111}) .
\end{equation}
The entanglement-dimensionality vector contains $2^{n-1}-1=63$ elements. When analyzing bipartitions $(j|\bar{j})$ with different numbers of particles in $j$ (that we label as $|j|$), we can apply our method to bound the elements of $\vec{v}$, and also assign them to partitions with different $|j|$. In other words, applying \cref{eq:obsMultipartite}, we can determine that the entanglement-dimensionality vector is $\vec{v}=(\vec{v}_{|j|=1} , \vec{v}_{|j|=2}, \vec{v}_{|j|=3})$, where the subvectors are $\vec{v}_{|j|=1}=(2_{6},1_{1}), \vec{v}_{|j|=2}=(4_{8},2_{12},1_{1}), \vec{v}_{|j|=3}=(4_{20},2_{14},1_{1})$. Here, the subscript indicates the number of times each element is repeated. The element $1$ in each subvector immediately implies that the state can be partitioned into three parties as $(1|23|4567)$. Thus, the state is $3$-separable. The entanglement depth is determined by the number of particles in the largest party, which in this case is 4.

Thus, from this example, we see how it is possible to characterize entanglement depth and $k$-separability from the entanglement-dimensionality vector in the case that the state is pure, which {clarifies} the entanglement structure of the state.
At the same time, the entanglement-dimensionality vector $\vec v$ provides additional information, namely the entanglement dimensionalities across all bipartitions, which is a deeper quantifier of entanglement, and can also lead to more quantitative conclusions, e.g., on the complexity of the state for simulation or on its applicability to specific tasks.

See SM for further details on the proof and this example~\cite{supplement}.

\bibliography{references}

\clearpage
\newpage
\setcounter{page}{1}
\renewcommand{\theequation}{S\arabic{equation}}
\setcounter{equation}{0}
\renewcommand{\thefigure}{S\arabic{figure}}
\setcounter{figure}{0}
\onecolumngrid

{\centering\Large\bfseries Supplemental Material: Uncertainty relations between quantum Fisher information and entanglement monotones\par}

\section{Proof of Observation 1 }\label{sec:ContinuousMonotoneProof}

First, we show how to convert the QFI matrix inequality into a scalar inequality, where we define the quantity $T^{(t)}(\varrho)$, which we show to be convex w.r.t the pure state decomposition of $\varrho$. In analogy with \cref{eq:QFIMBlockDef}, we define the following submatrices
\begin{equation}\label{eq:PureStateCMBlocksDef}
\left(\begin{array}{cc}
\kappa_a & X \\
X^T & \kappa_b
\end{array}\right):=
\sum_k p_k\Gamma_{\psi_k}(\vec g),
\end{equation}
\begin{equation}
\Delta=
\left(\begin{array}{cc}
\Delta_a & \Delta_X \\
\Delta_X^T & \Delta_b
\end{array}\right):=4\sum_k p_k\Gamma_{\psi_k}(\vec g)-\mathcal{F}_\varrho(\vec g).
\end{equation}
The latter should be positive for all quantum states with arbitrary pure-state decomposition $\varrho=\sum_kp_k\ket{\psi_k}\bra{\psi_k}$ according to the QFIM matrix inequality, which follows from the convexity of the QFIM. For clarity, let us also recall the definition of the block form of QFIM $\mathcal{F}_\varrho(\vec g)=\left(\begin{array}{cc}
\mathcal{F}_a & \mathcal X_\varrho \\
\mathcal X^T_\varrho & \mathcal F_b
\end{array}\right)$ (\cref{eq:QFIMBlockDef} in the main text).

Afterwards, let $O$ be the orthogonal matrix from the singular value decomposition of $\mathcal{X}_\varrho$, i.e., $\tr|\mathcal{X}_\varrho| = \tr(O^\top \mathcal{X}_\varrho)$. For $\alpha, \beta \in \mathbb R$, we construct the matrix
\begin{equation}
W = \left(\begin{array}{cc}
\alpha^2 I & \alpha\beta\, O \\
\alpha\beta\, O^\top & \beta^2 I
\end{array}\right)
\end{equation}
This matrix is positive semidefinite, as can be verified via the Schur complement: when $\beta \ne 0$, i.e. $\beta^2 I>0$, $W \geq 0$ if and only if $\alpha^2 I - \alpha^2 O O^\top \geq 0$, which holds since $O O^\top = I$. When $\beta = 0$, the matrix is also obviously positive semidefinite. Since $\Delta \geq 0$, we have
\begin{equation}
\tr\left(W \Delta\right) = \tr\left(W\left(4 \sum_k p_k \Gamma_{\psi_k} - \mathcal{F}_\varrho\right)\right) \geq 0.
\end{equation}
Expanding the two terms of this inequality, we obtain
\begin{equation}
\tr\left(W \mathcal{F}_\varrho\right) = \alpha^2 \tr(\mathcal{F}_a) + \beta^2 \tr(\mathcal{F}_b) + 2\alpha\beta\, \tr(O^\top \mathcal{X}_\varrho) = \alpha^2 \tr(\mathcal{F}_a) + \beta^2 \tr(\mathcal{F}_b) + 2\alpha\beta\, \tr|\mathcal{X}_\varrho|
\end{equation}
and
\begin{equation}
\tr\left(W \sum_k p_k \Gamma_{\psi_k}\right) = \alpha^2 \tr(\kappa_a) + \beta^2 \tr(\kappa_b) + 2\alpha\beta\, \tr(O^\top X),
\end{equation}
where the last term satisfies $\tr(O^\top X) \leq \tr|X|$. Combining these results yields
\begin{equation}\label{eq:obs1proof1}
\alpha^2 \tr(\mathcal F_a)+ \beta^2 \tr(\mathcal F_b) +2 \alpha \beta\, \tr|\mathcal X_\varrho | \leqslant 4 \left( \alpha^2 \tr(\kappa_a) + \beta^2 \tr(\kappa_b)+2 \alpha \beta \, \tr |X| \right) .
\end{equation}
In the cases that $\beta\ne 0$, by dividing \cref{eq:obs1proof1} by $\beta^2$, we obtain
\be
T^{(t)}(\varrho):=\tr(\mathcal F_b) + t^2 \tr(\mathcal F_a) + 2 t \ \tr|\mathcal X_\varrho | \leqslant 4 \left( t^2 \tr(\kappa_a) + \tr(\kappa_b)+2t \, \tr |X| \right)\le\sum_k p_k T^{(t)}(\psi_k),
\ee
where we called $t=\alpha/\beta$. Intuitively, $T^{(t)}(\varrho)$ represents a sum of QFIs associated to collective observables, which explains why it is expected to be convex. Set $t=1$ and we have
\begin{equation}
    T(\varrho)\le \sum _kp_kT(\psi_k).
\end{equation}
Furthermore, by substituting the trace norm with trace, we obtain~\cref{eq:TboundsumFGi} in the main text
\begin{equation}
    \begin{aligned}
        T(\varrho)&=\tr(\mathcal F_a) +\tr(\mathcal F_b) + 2 \ \tr|\mathcal X_\varrho |\ge \tr(\mathcal F_a) +\tr(\mathcal F_b) + 2 \ \tr\mathcal X_\varrho\\&=\sum_iF_\varrho (g_i^{(a)}\otimes \mathbb1)+\sum_iF_\varrho (\mathbb 1\otimes g_i^{(b)})+2 \ \tr\mathcal X_\varrho\\&= \sum_i F_\varrho(g_i^{(a)}\otimes \mathbb1+\mathbb 1\otimes g_i^{(b)}),
    \end{aligned}
\end{equation}where we used the relation between QFIM and QFI in the last equality.

We proceed to give the analytical expression of $T(\psi)$ given the Schmidt coefficients $\lambda_i$ of $\psi$. The marginal QFIs can be expressed compactly as
\begin{equation}
\begin{aligned}
    \tr(\mathcal F_a)
&=4\sum_k \Delta(g_k^{(a)}\otimes \id)^2_\psi
=4\sum_k [\langle (g_k^{(a)})^2\rangle_{\varrho_a} -\langle (g_k^{(a)})\rangle_{\varrho_a}^2]
\\&=4(d-\Tr \varrho_a^2)
=4\bigl[(d-1)+E_L(\psi)\bigr],
\end{aligned}
\end{equation}
where {$E_L(\psi) = 1 - \sum_i \lambda_i^2$} is the linear entanglement entropy of the pure state $\ket{\psi}$. The above derivation applies to $\tr(\mathcal F_b)$ as well and also appears in Proposition II.9 of Ref.\cite{gittsovich08}.
One can also find a simple closed‐form relation for the trace norm of the cross‐correlation block and entanglement monotones calculated on pure states:  
\begin{equation}
\frac14\tr\bigl|\mathcal X_\psi\bigr|
=\sum_i\left(\lambda_i-\lambda_i^2\right)
+2\sum_{i<j}\sqrt{\lambda_i\lambda_j}
=E_L(\psi)+2\mathcal N(\psi),
\end{equation}
where $\mathcal N(\psi) = (\| \ketbra{\psi}^{T_B} \| - 1 )/2 = \sum_{i<j}\sqrt{\lambda_i\lambda_j}$ is the negativity of $\ket{\psi}$. Therefore, $T^{(t)}(\psi)$ can be compactly written as\begin{equation}\label{pureT}
    T^{(t)}(\psi)=4(t^2+1)(d-1)+4(t+1)^2E_L(\psi)+16t\mathcal N(\psi),
\end{equation}where $E_L(\psi)=\frac12C^2(\psi)$, which also shows that $T^{(t)}(\varrho)$ is essentially an entanglement monotone. Alternatively, we may rewrite $T^{(t)}(\psi)$ as a function of marginal state $\varrho_A:=\tr_B(\ket{\psi}\bra{\psi})$:\begin{equation}
    T^{(t)}(\psi)=8E_T^{(t)}(\psi)+4(t^2+1)(d-1),
\end{equation}where we define\begin{equation}
    E_T^{(t)}(\psi)=\frac{(t^2+1)}2+t\, \tr^2(\sqrt{\varrho_a})-\frac{(t+1)^2}{2}\tr(\varrho_a^2).
\end{equation}For every non-negative $t$, it is concave with \textit{the reduced density matrix} $\varrho_a$ and satisfies the conditions proposed by \cite{vidal2000entanglement} of a legitimate pure-state entanglement monotone. Then, it can be extended to mixed states via the convex-roof construction:
\begin{equation}
    E_T^{(t)}(\varrho):=\inf_{D(\varrho)}\sum_k p_kE_T^{(t)}(\psi_k) .
\end{equation}
Note also that in the case $t=0$, this measure reduces to the convex-roof of the squared concurrence, i.e., the 2-tangle $\mathcal T(\varrho)$.

Now we try to develop a series of bounds to continuous entanglement monotones. The core of our method is based on the following chain inequalities of an arbitrary convex-roof entanglement monotone $G(\varrho)$:\begin{equation}\label{convexchain}
    G(\varrho)=\inf_{\mathcal{D}(\varrho)} \sum_k p_k G\left(\psi_k\right)=\sum_k p_k^{(o)}G(\psi_k^{(o)})\ge\sum_k p_k^{(o)}f(T(\psi_k^{(o)}))\ge  f(\sum_kp_k^{(o)}T(\psi_k^{(o)}))\ge f(T(\varrho)),
\end{equation}where we have used the optimal decomposition $\{p_k^{(o)},\ket{\psi_k^{(o)}}\}$ and assumed the function $f(x)$ to be convex and monotonically increasing, which justifies the second and the last inequality respectively. Specifically, since $T$ is convex, we have $\sum_k p_k^{(o)} T(\psi_k^{(o)}) \geq T(\varrho)$. The monotonicity of $f$ then implies $f\left(\sum_k p_k^{(o)} T(\psi_k^{(o)})\right) \geq f(T(\varrho))$. An immediate conclusion from this argument is obtained by considering $E_T^{(t)}$ and $f(x)=\frac{1}{8} x-\frac{1+t^2}{2}(d-1)$, which leads to the bound
\begin{equation}
    E_T^{(t)}(\varrho)\ge \frac18T^{(t)}(\varrho)-\frac{1+t^2}{2}(d-1).
\end{equation}
Here, we are essentially exploiting the fact that the convex-roof is the largest convex function that equals $E_T^{(t)}(\psi)$ for pure states, plus the fact that $T(\varrho)$ itself is convex. Because of that, it is noteworthy that this bound is expected to be tight. By letting $\beta=0$ or $\alpha=0$ in~\cref{eq:obs1proof1} respectively (corresponding to $t \to +\infty$ or $t=0$), the general bound reduces exactly to \cref{observation3} in terms of
\begin{equation}
\begin{aligned}
& \inf _{\mathcal{D}(\varrho)} 4 \operatorname{tr}\left(\kappa_a\right) \geqslant \operatorname{tr}\left(\mathcal{F}_a\right) ,\\
& \inf _{\mathcal{D}(\varrho)} 4 \operatorname{tr}\left(\kappa_b\right) \geqslant \operatorname{tr}\left(\mathcal{F}_b\right) ,
\end{aligned}
\end{equation}
where the left-hand side is connected to the 2-tangle.

For other continuous monotones, we may just focus on the pure-state cases, i.e. to find $E(\psi)\ge f(T(\psi))$ to finish our construction. For the concurrence, calling $\mathcal N(\psi)\le \sqrt{\frac{d(d-1)}{2}}C(\psi)$ and $C(\psi)\le \sqrt{\frac{2(d-1)}{d}}$, we derive the bound to pure-state concurrence\begin{equation}
    C(\psi)\ge \frac{\sqrt{d}(T(\psi)-8(d-1))}{4(d+2) \sqrt{2(d-1)}}
\end{equation}and further\begin{equation}
    C(\varrho)\ge \frac{\sqrt{d}(T(\varrho)-8(d-1))}{4(d+2) \sqrt{2(d-1)}}:=\mathcal B_C(\varrho).
\end{equation}

Now we use this bound to find a lower bound for the entanglement of formation, using the following argument.

Given a lower bound to the concurrence $\mathcal B_C(\psi)$ valid for pure states and a convex, monotonically increasing function $V(x)$ such that
$E_F(\psi) \geq V(C(\psi))$, we can get a lower bound to the entanglement of formation as
\begin{equation}
E_F(\varrho)=\sum_k p_k E_F\left(\psi_k\right) \geqslant \sum_k p_k V\left(C\left(\psi_k\right)\right)\geqslant V\left(\sum_k p_k C\left(\psi_k\right)\right) \geqslant V\left(\mathcal B_c\right) .
\end{equation}
In particular, we use the optimal such function, which was found in~\cite{ObservableZhang2011,DetectingZhang2013} and is given by
\begin{equation}
\label{eq:DefinitionCORL}
V(\lambda):=\operatorname{co}\left(R_L^{(d)}(\lambda)\right)= \begin{cases}H_2\left(\frac{1+\sqrt{1-\lambda^2}}{2}\right), & \lambda \in[0,1], \\ \frac{\log _2(k+1)-\log _2(k)}{\sqrt{\frac{2 k}{k+1}}-\sqrt{\frac{2 k-1}{k}}}\left(\lambda-\sqrt{\frac{2(k-1)}{k}}\right)+\log _2(k), & \lambda \in\left[1, \sqrt{\frac{2(d-1)}{d}}\right]\end{cases}
\end{equation}
where $k=\lfloor 2/(2-\lambda^2)\rfloor$. $H_2(x)=-x \log _2 x-(1-x) \log _2(1-x)$ is the standard binary entropy function and $\operatorname{co}(f(x))$ is the convex hull of $f(x)$.

\begin{figure}[h]
\centering
\includegraphics[width=0.6\columnwidth]{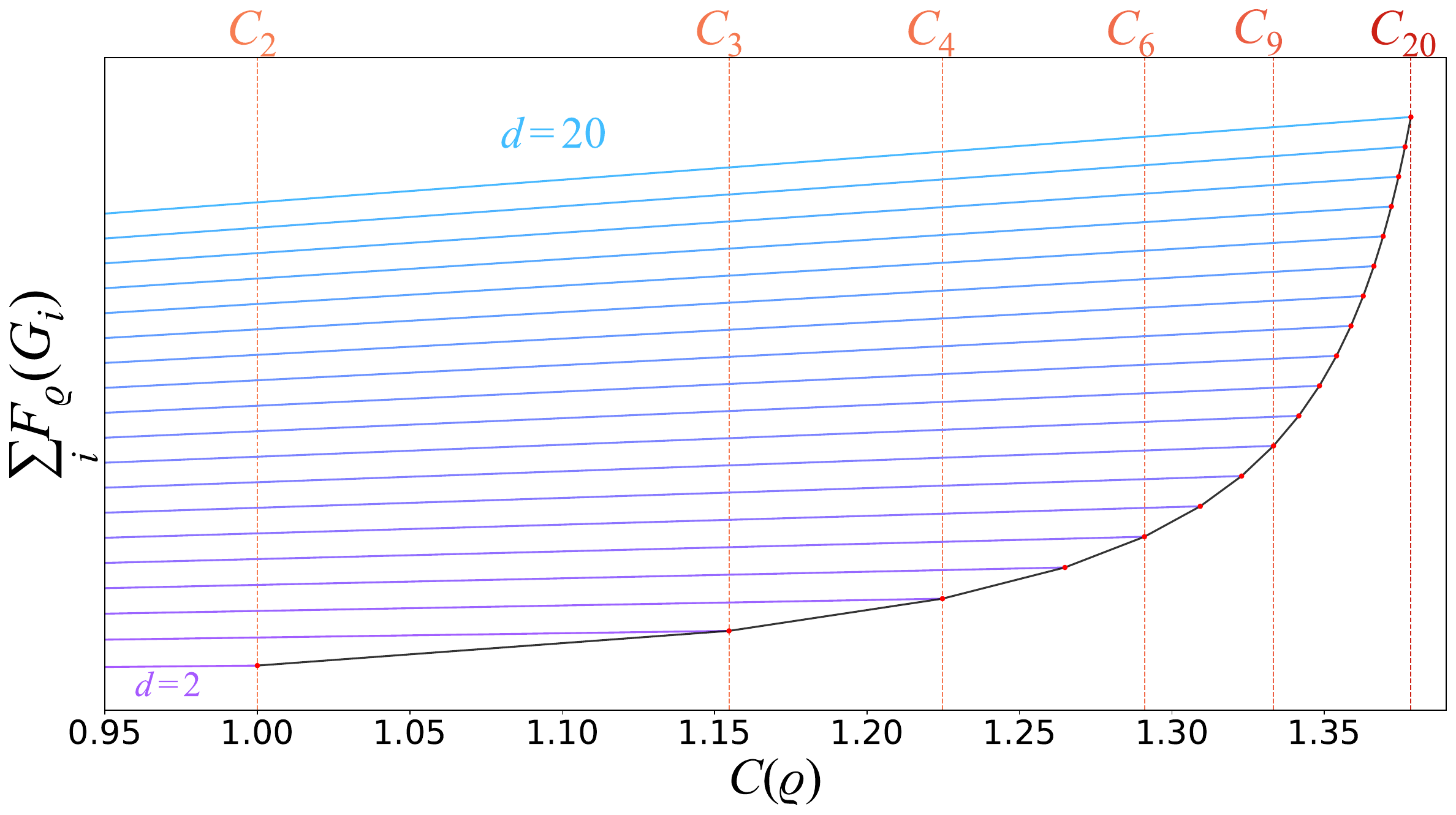}
\caption{Bounds on the sum of collective QFIs, $\sum_i F_\varrho(G_i)$, as functions of the concurrence $C(\varrho)$. We display bipartite high dimensional systems from $d=2$ to $d=20$. The purple-blue gradient curves represent the maximal value of $\sum_i F_\varrho(G_i)$ for each monotone, clearly showing that greater entanglement necessarily imposes a higher upper bound. It follows that states with only weak entanglement cannot achieve large $\sum_i F_\varrho(G_i)$. Red points on the solid black curves correspond to maximally entangled states (MES) of different dimensions. The dashed lines $C_i$ denote the corresponding entanglement monotone values of $i$-dimensional MES in the these figures as a reference. 
}
\label{fig:QFISumContinuousMonotones}
\end{figure}

See \cref{fig:QFISumContinuousMonotones} for a plot of the bound on the concurrence.

\section{An alternative proof of Observation 2 and examples}\label{sec:ObsAlternativeProof}

For the discrete Schmidt number, we can use the following relations
\begin{subequations}
  \begin{align}  
     \tr(\kappa_a) &=\sum_k p_k \sum_j \left( \av{(g_j^{(a)})^2}_{\psi_k} - \av{g_j^{(a)}}^2_{\psi_k} \right) = \sum_k p_k (d-\tr(\varrho^2_a)_k)\le d-\frac 1 r, \label{eq:boundtrkappaa} \\
         \tr(\kappa_b) &\le d-\frac 1 r, \\
    \tr|X_r|&= \tr|\sum_k p_k X_{\psi_k}|\le \sum_k p_k \tr|X_{\psi_k}| \le r-\frac 1 r , \label{eq:boundtrXaa}
  \end{align}
\end{subequations}
which are valid for the blocks of the boundary covariance matrix $\Gamma = \sum_k p_k \Gamma_{\psi_k}$ (blocks are defined in \cref{eq:PureStateCMBlocksDef}) where $\ket{\psi_k}$ are pure states with Schmidt rank smaller or equal than $r$.

Using those bounds and dividing \cref{eq:obs1proof1} by $\beta^2$ we obtain
\be
\tr(\mathcal F_b) + t^2 \tr(\mathcal F_a) + 2 t \ \tr|\mathcal X_\varrho | \leqslant 4 \left( (1+ t^2)\left( d - \tfrac 1 r \right) +2 t \left(r-\tfrac 1 r \right) \right) ,
\ee
where we called $t=\alpha/\beta$. 
Rearranging the terms, we get
\be\label{eq:Obs1ProofRearrangingWithT}
\left(\tr(\mathcal F_b) - 4\left( d - \tfrac 1 r \right) \right) + 2 t \left( \tr|\mathcal X_\varrho |  - 4 \left(r-\tfrac 1 r \right) \right) + t^2 \left( \tr(\mathcal F_a) - 4 \left( d - \tfrac 1 r \right)\right) \leq 0 , 
\ee
which should hold for all $t\in \mathbb R$. Furthermore, we have that for all states with Schmidt number smaller or equal than $r$, the following inequalities holds
\be
\begin{aligned}
    4 \left( d - \tfrac 1 r \right)- \tr(\mathcal F_a) & \geq 0 , \\
    4 \left( d - \tfrac 1 r \right)- \tr(\mathcal F_b) & \geq 0 ,
\end{aligned}
\ee
due to the positivity of the matrix $\Delta := 4 \Gamma_r - \mathcal F_\varrho(\vec g)$, and its principal minors $\Delta_a$ and $\Delta_b$.

Maximizing the left-hand side over all $t$ we get
\be
-\left( 4 \left( d - \tfrac 1 r \right)- \tr(\mathcal F_a) \right) \left( 4 \left( d - \tfrac 1 r \right)- \tr(\mathcal F_b) \right) + \left( \tr|\mathcal X_\varrho |  - 4 \left(r-\tfrac 1 r \right) \right)^2 \leq 0 
\ee
which is obtained for  $t=\left( \tr|\mathcal X_\varrho |  - 4 \left(r-\tfrac 1 r\right) \right)\bigg/ \left( 4 \left( d - \tfrac 1 r \right)- \tr(\mathcal F_a) \right)$ that must have a positive denominator for all states with Schmidt number bounded by $r$.

Thus, in total we conclude that all states $\varrho$ with a Schmidt number upper bounded by a given $r$ must satisfy the three following inequalities
\be
\begin{aligned}
     \tr|\mathcal X_\varrho | &\leq \sqrt{\left( 4 \left( d - \tfrac 1 r \right)- \tr(\mathcal F_a) \right) \left( 4 \left( d - \tfrac 1 r \right)- \tr(\mathcal F_b) \right)} + 4 \left(r-\tfrac 1 r \right) , \\
    \tr(\mathcal F_a) & \leq 4 \left( d - \tfrac 1 r \right) , \\
    \tr(\mathcal F_b) & \leq 4 \left( d - \tfrac 1 r \right) ,
\end{aligned}
\ee
which completes the proof. 

Note also that in the above proof, we have optimized over the real number $t$. However, for each real value of $t$, one obtains, in principle, a valid Schmidt number condition. One can also deal with the other terms to get weaker bounds, but in terms of fewer quantities.For example, a different way of finding \cref{eq:FisherTraceSum} is to substitute $t=1$ in \cref{eq:Obs1ProofRearrangingWithT}. This way, we get $\tr(\mathcal F_b) +  \tr(\mathcal F_a) + 2\tr|\mathcal X_\varrho |  \leq  8\left( d + r - \tfrac 2 r \right)$, which is a stronger version of \cref{eq:FisherTraceSum}. The latter is obtained from the equation above by further bounding the left-hand side from below using $\tr(\mathcal X_\varrho) \leq \tr|\mathcal X_\varrho|$. This condition contains the sum of the QFIs for a full collective basis of the form $G_k = g_k^{(a)} \otimes \id + \id \otimes g_k^{(b)}$ and clearly depends on the chosen bases. When such bases are optimally chosen, this criterion will be equivalent to \cref{eq:FXTraceNormUpperBound}, supposing $\tr(\mathcal F_a)=\tr(\mathcal F_b)$. 

\begin{figure}[h]
\centering
\includegraphics[width=0.5\columnwidth]{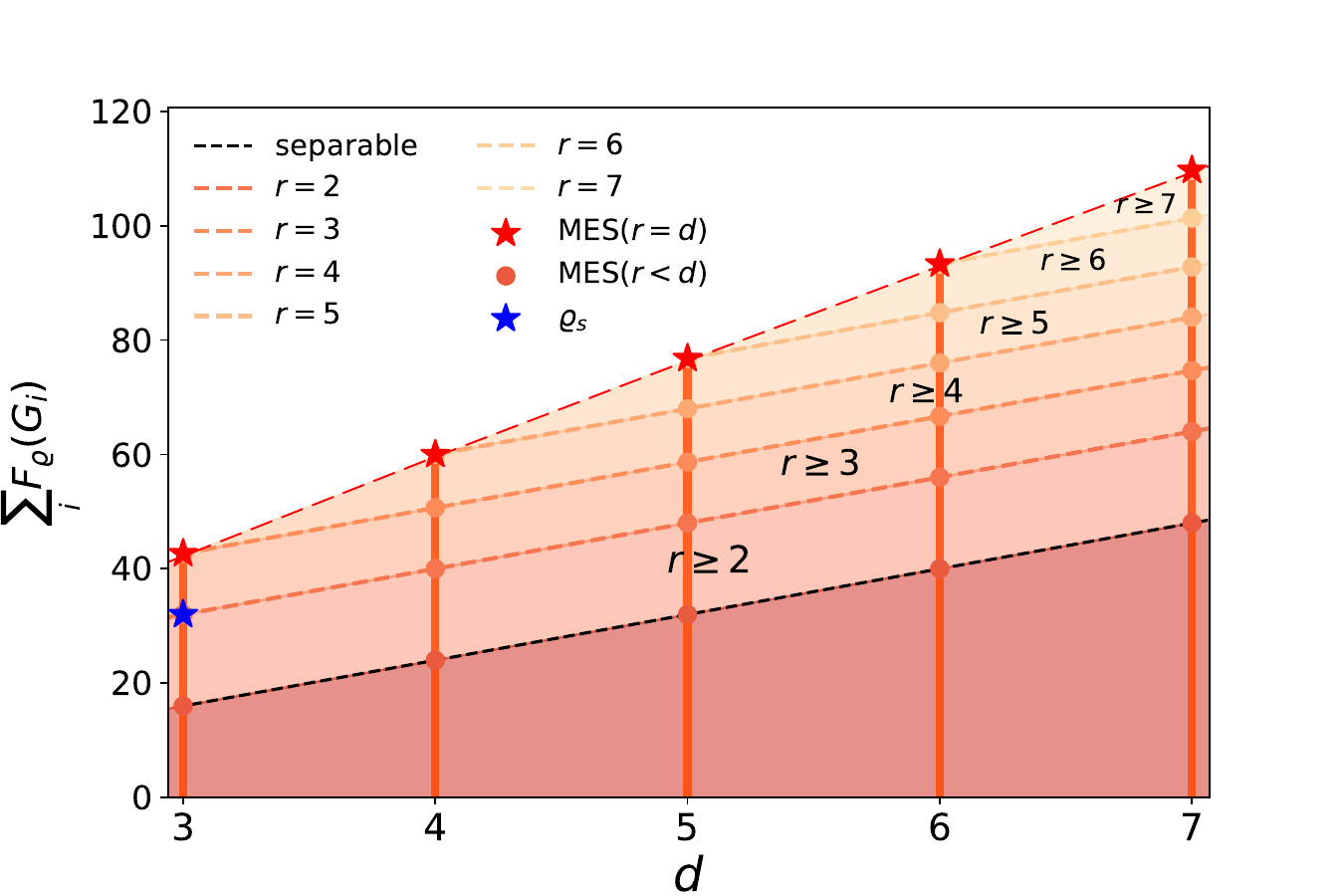}
\caption{Bounds on the sum of collective QFIs, $\sum_i F_\varrho(G_i)$, for different entanglement dimensionalities. The vertical orange lines are the attainable regions for all physical bipartite quantum states with different dimensions. The dashed lines are the upper bounds of $\sum_i F_\varrho(G_i)$ for states with a given Schmidt number $r$, where the points and red stars on these lines correspond to maximally entangled states (MES) $\frac{1}{\sqrt{r}} \sum_{i=0}^{r-1} \ket{ii}$. The blue star marks a special mixed state $\varrho_s$ that reaches the bound for $d=3$ and $r=2$ (see main text). Larger sums of QFIs are achieved with higher dimensions. 
}
\label{fig:FQAndCovVSR}
\end{figure}

See~\cref{fig:FQAndCovVSR} for a plot of the bounds.

We note that the bound in \cref{eq:FXTraceNormUpperBound} in the main text is not only saturated by Schmidt-rank-$r$ maximally entangled pure states, but also by certain mixed states. For example, for $d=3$ and $r=2$, it is also saturated by the mixed state $\varrho_s=\sum_{i=1}^3 p_i \ketbra{\psi_i}$, which is an arbitrary mixture of $\ket{\psi_1}=\frac1{\sqrt{2}}(\ket{00}-\ket{11})$, $\ket{\psi_2}=\frac1{\sqrt{2}}(\ket{12}+\ket{20})$, and $\ket{\psi_3}=\frac1{\sqrt{2}}(\ket{02}+\ket{21})$. 
Furthermore, when the probabilities satisfy $0\leq p_i\leq 0.5$, the state $\varrho_s$ is $2$-unfaithful according to the definition given in~\cite{WeilenmannPRL2020BeyondFidelity}, which means that its entanglement is undetectable by any fidelity witness. 

\section{Proof of no-go theorem of Schmidt number detection by single QFI}\label{sec:NoGoTheoremProof}

The quantum Fisher information of an observable $M$ on a pure state $\ket{\psi}$ is given by 
\begin{equation}
    F_\psi(M)=4(\Delta M)_\psi^2 .
\end{equation}
For a mixed state $\varrho = \sum_k p_k \ketbra{\psi_k}$, the QFI is convex {so that we obtain}
\begin{equation}
    F_\varrho(M) \le 4\sum_k p_k (\Delta M)_{\psi_k}^2.
\end{equation}
Thus, in order to bound the QFI of an observable $M$ from above, it is sufficient to find an upper bound on the {\it variance} of $M$ valid for all pure states.

Let us now consider $M=A\otimes \mathbb 1+\mathbb 1\otimes B$ with $A,B$ being arbitrary local observables. 
For a pure product state $\ket{\psi} = \ket{\psi_a}\ket{\psi_b}$, we can make use of the additivity property and get:
\be
\var{M}_\psi = \var{A}_{\psi_a} + \var{B}_{\psi_b} ,
\ee
which is in turn easily observed to be upper-bounded by
\be
\var{A}_{\psi_a} + \var{B}_{\psi_b} \leq \frac14\left(\mu_{\max}(A) - \mu_{\min}(A) \right)^2 + \frac14\left(\mu_{\max}(B) - \mu_{\min}(B) \right)^2 ,
\ee
where we denoted by $\mu_{\max}(\cdot)$ and $\mu_{\min}(\cdot)$ the largest and smallest eigenvalues of an operator respectively. This is because the variance of an observable on a quantum state is upper-bounded by
\be
\var{A}_{\psi_a} \leq \frac14\left(\mu_{\max}(A) - \mu_{\min}(A) \right)^2 ,
\ee
with the state saturating this relation given by
\be
\ket{\psi_a}=\frac1{\sqrt 2} \left(\ket{\mu_{\max}(A)}+\ket{\mu_{\min}(A)}\right) .
\ee

This way, we find that the QFI of a collective observable $M$ is upper-bounded by
\be\label{eq:QFIboundr1}
F_\varrho(M) \le  \left(\mu_{\max}(A) - \mu_{\min}(A) \right)^2 + \left(\mu_{\max}(B) - \mu_{\min}(B) \right)^2  ,
\ee
for all {\it separable} states $\varrho$, i.e., for all states of Schmidt-number $r=1$. A state that saturates the upper bound is given by
\be
\ket \psi=\frac1{2} \left(\ket{\mu_{\max}(A)}+\ket{\mu_{\min}(A)}\right) \otimes \left(\ket{\mu_{\max}(B)}+\ket{\mu_{\min}(B)}\right) ,
\ee
where $\ket{\mu_{\max}(\cdot)}$ and $\ket{\mu_{\min}(\cdot)}$ denote eigenvectors corresponding to the maximal and minimal eigenvalues of the observables respectively. Note that if there are multiple such vectors, one can take any linear combination of them as $\ket{\mu_{\max}(\cdot)}$ and $\ket{\mu_{\min}(\cdot)}$.
  
On the other hand, the overall upper-bound on $F_\varrho(M)$, valid {\it for all quantum states}, can also be found easily with a similar reasoning and we get
\be
F_\varrho(M) \leq \left(\mu_{\max}(M) - \mu_{\min}(M) \right)^2 ,
\ee
which is in general strictly larger than the right-hand side of \cref{eq:QFIboundr1}. Note that $\mu_{\max}(M)=\mu_{\max}(A)+\mu_{\max}(B)$ and the same equality holds for its minimum. Therefore, the following inequality holds {\it for all quantum states}
\be\label{eq:FMBoundSchmidtrank2}
F_\varrho(M) \leq  \left(\mu_{\max}(A)+\mu_{\max}(B) - \mu_{\min}(A) - \mu_{\min}(B) \right)^2 ,
\ee
and is saturated by a state of the form
\be\label{eq:statesaturatingsingleFM}
\ket{\psi} = \frac 1 {\sqrt{2}} \left(\ket{\mu_{\max}(A)}\ket{\mu_{\max}(B)}+\ket{\mu_{\min}(A)}\ket{\mu_{\min}(B)} \right) ,
\ee
where again $\ket{\mu_{\max}(\cdot)}$ and $\ket{\mu_{\min}(\cdot)}$ are any of the eigenvectors corresponding to the maximal and minimal eigenvalues respectively. Any two vectors $\ket{\mu_{\min}(\cdot)}$ and $\ket{\mu_{\max}(\cdot)}$ are always orthogonal, since they correspond to different eigenvalues, and thus we can see that the state in \cref{eq:statesaturatingsingleFM} has Schmidt rank equal to two. Hence, the bound in \cref{eq:FMBoundSchmidtrank2} is valid for all states with Schmidt number $r\geq 2$.

Now consider an illustrative example from Ref.~\cite{HiguchiHow2000}: a 4-qubit pure state with high-dimensional bipartite entanglement and vanishing single collective-spin QFI \cite{HyllusFisher2012,TothMultipartite2012}
\begin{equation}\label{eq:PhiDef}
\ket{\Phi}
=\frac{1}{\sqrt{6}}\Big(
\ket{0011}+\ket{1100}
+\omega(\ket{1010}+\ket{0101})
+\omega^{2}(\ket{1001}+\ket{0110})
\Big),
\end{equation}
where {$\omega:=e^{\frac{2\pi {\rm i}}{3}}$}. Let $J_k=\tfrac12\sum_{\ell=1}^4 \sigma_k^{(\ell)}$ $(k=x,y,z)$ be the total spin operators. Then $F_{\Phi}(J_k)=4(\Delta J_k)^2_{\Phi}=0$ for $k=x,y,z$. This can be verified by expanding $\ket{\Phi}$ as a coherent superposition of three pairwise singlet products,
\begin{equation}\label{eq:PhiBellFactorize}
|\Phi\rangle=\frac{\rm i}{\sqrt{2}}\left|\psi^{-}\right\rangle_{12}\left|\psi^{-}\right\rangle_{34}+\frac{1}{\sqrt{6}}\left|\psi^{-}\right\rangle_{13}\left|\psi^{-}\right\rangle_{24}+\frac{1}{\sqrt{6}}\left|\psi^{-}\right\rangle_{14}\left|\psi^{-}\right\rangle_{23}
\end{equation}
where $\left|\psi^{-}\right\rangle=\frac{1}{\sqrt{2}}(|01\rangle-|10\rangle)$, each of which is annihilated by every $J_k$. Consequently, any single-$J_k$-based criterion, including the $k$-producible upper bounds of~\cite{HyllusFisher2012,TothMultipartite2012}, returns a trivial result ($F_Q=0$) and cannot detect any entanglement in $\ket{\Phi}$. If we consider a more general form of a collective operator, namely $\tilde J=\frac12\sum_{\ell=1}^{4} \sigma_{\hat n _\ell}^{(\ell)}$, entanglement can be detected via a single QFI. Although the solution of this problem requires numerical analysis~\cite{HyllusFisher2012}, we showcase this with an illustrative example by choosing $\tilde J=\frac12(\sigma_x^{(1)}+\sigma_x^{(2)}-\sigma_x^{(3)}-\sigma_x^{(4)})$. The QFI of $\tilde J$ is $\frac{16}3$ so that entanglement depth of $2$ is certified. However, any such more general single QFI criterion still falls in the class included in our no-go theorem described above. Thus, the dimensionality of entanglement of the state cannot be certified with any such $\tilde J$.

Nevertheless, $\ket{\Phi}$ is genuinely 4-partite entangled (entangled across every bipartition, e.g. composed of $1$ vs $3$ particles) and exhibits Schmidt rank $4$ across all bipartitions of $2$ vs $2$ particles. For instance, across the $(12|34)$ partition, the Schmidt coefficients are $\lambda=(\tfrac12,\tfrac16,\tfrac16,\tfrac16)$, and the same holds for the $(13|24)$ and the $(14|23)$ bipartitions. This genuinely high-dimensional bipartite entanglement contrasts with a $4$-qubit GHZ state, which has entanglement depth $4$ but only Schmidt rank $2$ across any bipartition.

Let us also calculate the entanglement of formation of this state across the $2$ vs $2$ bipartitions. For pure states, the entanglement of formation across a bipartition equals the Shannon entropy of the corresponding Schmidt coefficients. Thus, from $\lambda=(\tfrac12,\tfrac16,\tfrac16,\tfrac16)$, we obtain
\begin{equation}\label{eq:EoFPurePhi}
E_F(\ket{\Phi})=-\sum_i \lambda_i\log_2\lambda_i
=\tfrac12+\tfrac12\log_2 6\ \approx\ 1.7925
\end{equation}

Treating $\ket{\Phi}$ as a pure bipartite state of local dimension $d=4$ under a $2$ vs $2$ bipartition, our QFIM-based criterion yields $\mathcal{SN}(\ket{\Phi})\ge4$, correctly detecting $4$-dimensional  entanglement across all $2$ vs $2$ bipartitions. Additionally, our EoF criterion in \cref{eq:EoFBound} gives $E_F(\ket{\Phi}) \geq 1.4025$, which, in particular, exceeds the Bell-state value ($1$ ebit).

\section{Criteria based on the QFIs of collective spin operators}\label{sec:CollectiveSpin}

Let us first consider a generic set of collective observables $\{M_k\}=\{A_k \otimes \id + \id \otimes B_k\}$. Expressing a generic Schmidt-rank-$r$ state as $\ket{\psi_r}=\sum_{k=1}^r \sqrt{\lambda_k} \ket{u_k, v_k}$, we have to solve the following problem
\begin{equation}
\begin{aligned}
\mathcal B_r(\vec M) &= 4\max_{\lambda_1,\dots , \lambda_r} \sum_k \left[\left(\sum_{l,m=1}^r \sqrt{\lambda_l \lambda_m} \bra{u_l, v_l} M^2_k \ket{u_m, v_m} \right) \right. \\
&\left. - \left(\sum_{l,m=1}^r \sqrt{\lambda_l \lambda_m} \bra{u_l, v_l} M_k \ket{u_m, v_m} \right)^2 \right].
\end{aligned}
\end{equation}
As a paradigmatic example, let us consider the spin operators in the spin-$j$ representation where $j=(d-1)/2$ for a particle with $d$ levels. Those operators are defined as
\begin{equation}
\begin{aligned}
        j_x &=\frac12\sum_{l<k}\delta_{l+1,k}\sqrt{j(j+1)-m_l(m_l+1)}(\ket k\bra l+\ket l\bra k), \\
        j_y &=\frac i2\sum_{l>k}\delta_{l,k+1}\sqrt{j(j+1)-m_k(m_k+1)}(\ket k\bra l-\ket l\bra k), \\
        j_z &=\sum_k m_k\ket k\bra k ,
\end{aligned}
\end{equation}
with $-j\le m_k\le j$. The collective spin operators are $J_k^\pm=j_k\otimes \mathbb{1}\pm \mathbb{1}\otimes j_k, k=x,y,z$.

We can then consider upper bounds of the form
\be
\begin{aligned}
\max_{\mathcal{SN}(\psi_r)=r} \sum_{k=x,y,z} (\Delta J_k^\pm)_{\psi_r}^2 &:= \mathcal B_r(\vec J^\pm) , \\
\max_{\mathcal{SN}(\psi_r)=r} \sum_{k=x,y} (\Delta J_k^\pm)_{\psi_r}^2 &:= \mathcal B_r(\vec J_\perp^\pm) , 
\end{aligned}
\ee
and numerically maximize over all states with a given Schmidt rank. As a concrete example, we consider a $4$-dimensional system and find the values of the bounds above, which are listed in Table~\eqref{eq:valuesBJs} below
\begin{equation}\label{eq:valuesBJs}
\begin{array}{|c|c|}
\hline & \mathcal B_r(\vec J_\perp^\pm) \\
\hline r=1  & 7 \\
\hline r=2  & 11 \\
\hline r=3  & 11.5601 \\
\hline r=4  & 12 \\
\hline
\end{array} \qquad \qquad
\begin{array}{|c|c|}
\hline & \mathcal B_r(\vec J^\pm) \\
\hline r=1 & 7.5 \\
\hline r=2 & 12 \\
\hline r=3 & 13.3403 \\
\hline r=4 & 15 \\
\hline
\end{array}
\end{equation}
In order to investigate those bounds analytically, we follow a similar argument as in \cite{Liu2024bounding}, and consider Schmidt rank-$r$ states such that their Schmidt bases are aligned with the $j_z$ eigenbases. Numerical analysis also suggests that this is often the optimal choice, but so far a fully analytical optimization remains an open question. If we assume that the Schmidt bases of the optimal state $\ket{\psi_r}$ are aligned with the $j_z$ eigenbases, the upper bound on the sum of collective variances can be written as
\begin{equation}
    \Delta^2 J_x^\pm+\Delta^2 J_y^\pm\le \max_{\vec \lambda}\sum_{k=1}^{d-1} 2w_k^2(\sqrt{\lambda_k}+\sqrt{\lambda_{k+1}})^2,
\end{equation}
where we called $w_k:=\frac1{\sqrt2}\sqrt{j(j+1)-m_k(m_k+1)}$ and for simplicity, we have chosen $m_k=k-1-j$. In this case, the upper bound of $r=2$ can be solved analytically and one obtains 
\begin{equation}
\begin{aligned}
        \Delta^2 J_x^\pm+\Delta^2 J_y^\pm &\le 2 j^2+2 j-1+\sqrt{j^4+2 j^3+j^2+1} \;\; \text{($j$ integer)} , \\
        \Delta^2 J_x^\pm+\Delta^2 J_y^\pm &\le -\frac14 +3j(j+1) \quad \text{($j$ half-integer)} ,
\end{aligned}
\end{equation}
while the global upper bound valid for all states is given by
\begin{equation}
    \Delta^2 J_x^\pm+\Delta^2 J_y^\pm\le 2j(2j+1),
\end{equation}
which is saturated by some full rank states which are not maximally entangled. Similarly, the upper bound for the sum of three variances reads
\begin{equation}
\begin{aligned}
&\Delta^2 J_x^\pm+\Delta^2 J_y^\pm +\Delta^2 J_z^\pm\\
&\le \max_{\vec \lambda}\sum_{k=1}^{d-1} 2w_k^2(\sqrt{\lambda_k}+\sqrt{\lambda_{k+1}})^2+4\sum_{i=1}^d\lambda_i m_i^2-4\left(\sum_{i=1}^d\lambda_i m_i\right)^2,
\end{aligned}
\end{equation}
which can be easily solved numerically.

\section{Certification of multipartite entanglement structure using QFIM}\label{sec:MultipartiteEntanglement}

In the multipartite case, one has to look at the vector of entanglement entropies across all bipartitions, $\vec{\mathcal{E}}(\varrho)$, which is also further ordered non-increasingly~\cite{HuberStructure2013}. For an $n$-particle system, the number of bipartitions is $\mathcal{P}=2^{n-1}-1$, which is the number of components of $\vec{\mathcal{E}}(\varrho)$, and each such component is difficult to calculate in general, even with full knowledge of the density matrix. Current methods for detecting multipartite high-dimensional entanglement include fidelity witnesses~\cite{malik2016multi,erhard2018experimental,CerveraExperimental2022,hu2020experimental,bao2023very}, linear entropy vector~\cite{HuberStructure2013}, correlation tensor norm~\cite{klockl2015characterizing}, and also covariances~\cite{NonlinearLiuArxiv2024}, building upon the covariance matrix criterion developed in \cite{Liu2024bounding}. This latter approach can be also adapted to exploit the method presented here, in particular \cref{eq:FXTraceNormUpperBound}, to find a relationship between QFIM and multipartite high-dimensional entanglement.

Concretely, in a multipartite setting the so-called entanglement entropy vector is defined element-wise as $\mathcal{E}_j^\downarrow(\varrho):=\inf _{\mathcal{D}(\varrho)} \sum_k p_k \mathcal{E}_j^\downarrow\left(\psi_k\right)$, labeled with an index scanning all bipartitions $j\in \{1,2\dots , \mathcal P\}$ ~\cite{HuberStructure2013}, where $\mathcal{D}(\varrho)$ is a pure state decomposition $\varrho=\sum_k p_k \ket{\psi_k}\bra{\psi_k}$, and $\mathcal{E}_j^\downarrow\left(\psi_k\right)$ denotes the $j$-th largest entanglement entropy of $\ket{\psi_k}$ across all bipartitions. Thus, for an optimal decomposition $\varrho=\sum_k p_k \ket{\phi_k}\bra{\phi_k}$, the relation $\mathcal{E}^\downarrow_j(\phi_k) \leq \mathcal{E}^\downarrow_j(\varrho)$ holds for all $j$.

We showcase how zero-entropy (i.e., the Schmidt number) can be used to extend \cref{eq:FXTraceNormUpperBound} from the bipartite to the multipartite case. Using this approach, we obtain a bound on the entanglement-dimensionality vector $\mathcal{SN}^\downarrow_j(\varrho)=\inf_{\mathcal{D}(\varrho)} \max_{\ket{\psi_k} \in \mathcal{D}(\varrho)} \mathcal{SN}^\downarrow_j(\psi_k)$, where the notation is as defined above, and show that this bound can be computed via a linear program:

\setcounter{observation}{3}
\begin{observation}
For an $n$-partite state $\varrho$, consider a candidate entanglement-dimensionality vector $\vec v=(r_1 , r_2 , \dots , r_{\mathcal P})$. $\vec v$ is feasible for $\varrho$ if there exist a pure-state decomposition $\{p_k,\left|\psi_k\right\rangle\}$ and integer vectors $\vec r_k = ((r_1)_k,(r_2)_k, \dots ,(r_{\mathcal P})_k)$, together with permutation matrices $M_k$, such that $\vec r_k \leq M_k \vec v$ element-wise for all $k$ and
    \be\label{eq:obsMultipartiteSM}
     \left\{ \begin{array}{cc}
            h_{1}(\varrho) &\leq \sum_k p_k\left((r_{1})_k-\frac{2}{(r_{1})_k}\right) , \\   
            h_2(\varrho) &\leq \sum_k p_k\left((r_{2})_k-\frac{2}{(r_{2})_k}\right) , \\   
             \vdots \\ 
            h_{\mathcal P}(\varrho) &\leq \sum_k p_k\left((r_{\mathcal P})_k-\frac{2}{(r_{\mathcal P})_k}\right),  
    \end{array}
    \right.
    \ee
where the functions $h_j(\varrho)$ are defined in \cref{eq:SNVCritSM} for the bipartition $(j|\bar j)$. Finally, taking the element-wise minimum over all feasible vectors, the resulting vector $\vec v^{\min}=(v_1^{\min},\dots,v_{\mathcal P}^{\min})$ is the detected entanglement-dimensionality vector.
\end{observation}

\textbf{Proof.---}Consider a quantum state $\varrho$ composed of $n$ particles of dimension $d$. To certify its entanglement-dimensionality vector, we consider the QFIM for each bipartition $(j | \bar{j})$, labelled by a integer index 
$j \in \{1,2\dots , \mathcal P\}$: 
\begin{equation}
\mathcal{F}_\varrho(\vec g_j)=\left(
\begin{array}{cc}
\mathcal F_j & \mathcal X_\varrho^{(j|\bar{j})} \\
(\mathcal X_\varrho^{(j|\bar{j})})^T & \mathcal F_{\bar{j}} 
\end{array}
\right) ,
\end{equation}
where $\vec g_j$ is a basis pair for the partition $(j|\bar j)$. One example of such basis is obtained by the tensor product of the single-particle $su(d)$ bases for all particles in each party. 

Now, let us recall that a state $\varrho$ that has a given entanglement-dimensionality vector $\vec v$ can be decomposed by definition as
\begin{equation}
\varrho=\sum_k p_k \ket{\psi_k}\bra{\psi_k} ,
\end{equation}
where each pure state $\ket{\psi_k}$ has a Schmidt rank vector $\vec r_k$ that is such that $\vec r_k \leq M \vec v$ for some permutation matrix $M$.

Due to the convexity of QFIM, we obtain
\be
\mathcal{F}_\varrho(\vec g_j) \leqslant \sum_k p_k \mathcal{F}_{\psi_k}(\vec g_j) = 4 \sum_k p_k \Gamma_{\psi_k}(\vec g_j) , 
\ee
and by multiplying both sides by the vector $\boldsymbol{t}_\mu=\left(A \boldsymbol{u}_\mu,B \boldsymbol{v}_\mu\right)^T$ and its transpose, we arrive at an inequality similar to \cref{eq:obs1proof1}, namely
\begin{equation}
\begin{aligned}
 A^2 \tr(\mathcal F_j ) +B^2 \tr (\mathcal F_{\bar j})  +2 A B\ \tr|\mathcal X_\varrho^{(j|\bar{j})} | 
\leqslant & 4 A^2 \sum_k p_k\left(d-\frac{1}{(r_j)_k}\right)+4 B^2 \sum_k p_k\left(d-\frac{1}{(r_j)_k}\right)+8 A B \sum_k p_k\left((r_j)_k-\frac{1}{(r_j)_k}\right) ,
\end{aligned}
\end{equation}
where we also used the bounds in \cref{eq:boundtrkappaa,eq:boundtrXaa}.
After optimizing $A$ and $B$ over real values, we obtain
\begin{equation}
\frac{1}{4}\tr|\mathcal X_\varrho^{(j|\bar{j})} |-\sum_k p_k\left((r_j)_k-\frac{1}{(r_j)_k}\right) \leqslant \sqrt{\left(\sum_k p_k\left(d-\frac{1}{(r_j)_k}\right)-\frac{1}{4} \tr(\mathcal F_j )\right)\left(\sum_k p_k\left(d-\frac{1}{(r_j)_k}\right)-\frac{1}{4} \tr(\mathcal F_{\bar j})\right)} .
\end{equation}
Then, using the condition $\sqrt{(x-t)(y-t)}\leqslant \sqrt{x y}-t$, we arrive at
\begin{equation}\label{eq:SNVCritSM}
h_{j}(\varrho):= \frac{1}{4}\tr|\mathcal X_\varrho^{(j|\bar{j})} | -\sqrt{\left(d-\frac{1}{4} \tr(\mathcal F_{j})\right)\left(d-\frac{1}{4} \tr(\mathcal F_{\bar j})\right)} \leqslant \sum_k p_k\left((r_j)_k-\frac{2}{(r_j)_k}\right) .
\end{equation}
Once again, the constraint to the right-hand side is that the vector of Schmidt ranks across all bipartitions of $\ket{\psi_k}$ must satisfy $\vec r_k \leq M \vec v$ for some permutation matrix $M$. Thus, we have to check simultaneously the system of inequalities
    \be\label{eq:obsMultipartiteSMapp}
     \left\{ \begin{array}{cc}
            h_{1}(\varrho) &\leq \sum_k p_k\left((r_{1})_k-\frac{2}{(r_{1})_k}\right) , \\   
            h_2(\varrho) &\leq \sum_k p_k\left((r_{2})_k-\frac{2}{(r_{2})_k}\right) , \\   
             \vdots \\ %\quad &\quad   \\
            h_{\mathcal P}(\varrho) &\leq \sum_k p_k\left((r_{\mathcal P})_k-\frac{2}{(r_{\mathcal P})_k}\right),  
    \end{array}
    \right. .
    \ee
We denote by $\{\vec{v}\}_{\text{feas}}$ the set of all non-increasingly ordered vectors $\vec v$ such that the inequalities in \cref{eq:obsMultipartiteSMapp} are all satisfied. By definition of $\mathcal{SN}^\downarrow(\varrho)$, we obtain a rigorous lower bound to the entanglement-dimensionality vector by taking the element-wise minimum over all feasible vectors,
\be
v_j^{\min}:=\min_{\vec v\in \{\vec{v}\}_{\text{feas}}} v_j .
\ee
This vector $\vec v^{\min}$ satisfies $\mathcal{SN}^\downarrow_j(\varrho)\ge v_j^{\min}$ for all $j$.
\qed

To facilitate comprehension, we explain below the derivation of the full entanglement-dimensionality vector for the state in \cref{eq:7qubitstateapp}, which we report here as well for clarity:
\begin{equation}\label{eq:7qubitstateapp}
\ket{\phi}=\frac{1}{2\sqrt{2}}(\ket{0}+\ket{1})\otimes(\ket{00}+\ket{11})\otimes(\ket{0000}+\ket{1111}) .
\end{equation}
We recall that its entanglement-dimensionality vector is given by
\begin{equation}\label{eq:SNVPure7Qubit}
  \vec{v}=(\vec{v}_{|j|=1} , \vec{v}_{|j|=2} , \vec{v}_{|j|=3}) \quad \text{with} \quad \vec{v}_{|j|=1}=(2_{6},1_{1}),\quad \vec{v}_{|j|=2}=(4_{8},2_{12},1_{1}),\quad \vec{v}_{|j|=3}=(4_{20},2_{14},1_{1}) ,
\end{equation}
where the subscript indicates the number of times each element is repeated. 

We evaluate \cref{eq:obsMultipartiteSMapp} by inserting on the right-hand side the values $r-\frac{2}{r}=-1$ for $r=1$, $r-\frac{2}{r}=1$ for $r=2$ and $r-\frac{2}{r}=3.5$ for $r=4$ and calculating the quantities $h_j$ depending on the partition sizes, which results in
\begin{equation}
\vec{h}_{|j|=1}=(1_{6},-1_{1}),\quad \vec{h}_{|j|=2}=(3.5_{8},1_{12},-1_{1}),\quad \vec{h}_{|j|=3}=(3.5_{20},1_{14},-1_{1}),
\end{equation}
meaning that the entanglement-dimensionality vector is precisely that in \cref{eq:SNVPure7Qubit}.

For better clarity, we also summarize in the table below how to reconstruct its entanglement structure.

\begin{equation}\nonumber
\begin{array}{|c|c|c|}
\hline \text{Partition size} & \text {The number of elements equal to } 1 \text { in } \vec{v} & \text{Bipartition }(j|\bar{j}) \\
\hline|j|=1 & \geqslant 1 & 1 | 234567 \\
\hline|j|=2 & \geqslant 1 & 14567 | 23 \\
\hline|j|=3 & \geqslant 1 & 123 | 4567 \\
\hline
\end{array}
\end{equation}

Thus, for pure states, our method can allow to extract the full decomposition of the state, i.e., both the minimal number of parties and the size of each party.  For mixed states instead, it is in general not possible to deduce the optimal pure state decomposition that clearly shows its exact entanglement structure, but only the least dimensionalities needed to describe them. In that case, our method can thus be used to detect complementary information about the entanglement structure as compared to the depth and the number of partitions.

\section{Connection between the QFIM bound and quantum metrology}\label{sec:Lemma1Proof}

First, it is useful to introduce the following statement that follows from the multiparameter quantum Cram\'er-Rao bound (QCRB) :

\begin{lemma}\label{lemma1}
Given a set of parameters $\{\theta_i\}_{i=1}^{K}$, encoded in a quantum state $\varrho$ via a unitary evolution generated by a set of $\{\tilde H_i \}_{i=1}^K$, the following ``uncertainty relation'' holds:
\begin{equation}
    \left( \sum_{i=1}^K (\delta \theta_i)^2 \right) \cdot \left( \sum_{i=1}^K F_\varrho(\tilde H_i) \right) \ge K^2,
\end{equation}
{where the $(\delta \theta_i)^2$ are the variances of the estimators and  $K$ is the number of parameters.} 
\end{lemma}

\textbf{Proof. }Consider a single shot measurement, the QCRB reads 
\begin{equation}
    \Sigma \succeq \mathcal F_\varrho^{-1},
\end{equation}
where $\Sigma$ is the covariance matrix of the estimators of the parameters and $\mathcal F_\varrho = \mathcal F_\varrho(\vec{\tilde H})$ is the quantum Fisher information matrix (QFIM) associated to the estimation and we have assumed the non-trivial circumstance that the QFIM is of full rank so that the encoded parameters are independent. Taking the trace of this equation, we get
\be
\sum_{i=1}^K (\delta \theta_i)^2 = \tr(\Sigma) \geq \tr(\mathcal F_\varrho^{-1}) = \sum_i \vec n_i^T \mathcal F_\varrho^{-1} \vec n_i ,
\ee
where the vectors $\vec n_i$ form an orthonormal basis. Next, using that $\vec n^T M^{-1}\vec n\ge( \vec n^T M \vec n)^{-1}$ holds for any strictly positive definite matrix $M$ and unit vector $\vec n$, we obtain
\begin{equation}
    \sum_{i=1}^K (\delta \theta_i)^2  \ge \sum_i (\vec n_i^T \mathcal F_\varrho \vec n_i)^{-1}.
\end{equation}
Finally, by using the Cauchy-Schwartz inequality $(\sum_i  (\vec n_i^T \mathcal F_\varrho \vec n_i)^{-1}) (\sum_i \vec n_i^T \mathcal F_\varrho \vec n_i) \ge K^2$, we arrive at
\begin{equation}
    \sum_{i=1}^K (\delta \theta_i)^2  \ge \frac{K^2}{\sum_i \vec n_i^T \mathcal F_\varrho \vec n_i} =\frac{K^2}{\sum_{i=1}^K F_\varrho(\tilde H_i)} ,
\end{equation}
where in the last equation we used the fact that the diagonal elements of the QFIM are given by $F_\varrho(\tilde H_i)$.
\qed

Using \cref{lemma1}, we can now relate our criterion in \cref{eq:FisherTraceSum} (\cref{corollary1}) to a specific metrological task, in which $K=(d^2-1)$ parameters are encoded in a bipartite quantum state $\varrho$ by the unitary $U(\vec \theta)=\exp\left(-{\rm i}\sum_{k=2}^{d^2}\theta_k G_k\right)$, where $G_k = g^{(a)}_k\otimes \id + \id \otimes g^{(b)}_k$ are bipartite collective $su(d)$ generators. The task is to estimate the parameters in the proximity of zero, i.e., $\theta_i \simeq 0$, so that we have $\tilde H_i = G_i$. In this situation, the following bound holds for the variances of the estimators
\begin{equation}
    \sum_{i=2}^{d^2}\delta^2 \theta_i\ge \frac{(d^2-1)^2 r}{8(r^2+dr-2)} , 
\end{equation} 
and for all probe states that have a Schmidt number bounded from above by $r$, i.e., $\mathcal{SN}(\varrho)\leq r$. 

As an example, for $d=2$ with normalized Pauli matrices as $g_k$, entanglement is certified whenever $\sum_{k=x, y, z}\left(\delta \theta_k\right)^2<9 / 8$. 

Note that analogous reasoning could be made considering a smaller set of Hamiltonians. {We can also obtain} bounds for combinations of sensitivities through the QCRB and conditions of the form \eqref{eq:boundsforfewQFIs}. 

\end{document}